# Recovery of mechanical pressure in a gas of underdamped active dumbbells with Brownian noise

Marc Joyeux[(#)]

*LIPHY, Université Grenoble Alpes and CNRS, Grenoble, France*

**Abstract:** In contrast with a gas at thermodynamic equilibrium, the mean force exerted on a wall by a gas of active particles usually depends on the confining potential, thereby preventing a proper definition of mechanical pressure. In this paper, we investigate numerically the properties of a gas of underdamped self-propelled dumbbells subject to Brownian noise of increasing intensity, in order to understand how the notion of pressure is recovered as noise progressively masks the effects of self-propulsion and the system approaches thermodynamic equilibrium. The simulations performed for a mobile asymmetric wall separating two chambers containing an equal number of active dumbbells highlight some subtle and unexpected properties of the system. First, Brownian noise of moderate intensity is sufficient to let mean forces equilibrate for small values of the damping coefficient, while much stronger noise is required for larger values of the damping coefficient. Moreover, the displacement of the mean position of the wall upon increase of the intensity of the noise is not necessarily monotonous and may instead display changes of direction. Both facts actually reflect the existence of several mechanisms leading to the rupture of force balance, which tend to displace the mean position of the wall towards different directions and display different robustness against an increase of the intensity of Brownian noise. This work therefore provides a clear illustration of the fact that driving an autonomous system towards (or away from) thermodynamic equilibrium may not be a straightforward process, but may instead proceed through the variations of the relative weights of several conflicting mechanisms.

**Keywords:** Active matter, nonequilibrium pressure, underdamped dynamics, self-propelled dumbbells

**PACS**:   05.40.-a, 05.70.Ce, 02.70.Ns

[(#)] email : marc.joyeux@univ-grenoble-alpes.fr

## I. Introduction

Active matter defeats many intuitions based on the knowledge of the properties of systems at thermodynamic equilibrium. For example, the (mechanical) pressure of a gas at thermodynamic equilibrium can be measured as the mean force per unit area exerted by the constituent particles on a confining wall. Quite importantly, identical results are obtained for walls made of different materials, that is, the pressure does not depend on the precise interactions between the particles and the wall. Moreover, the (thermodynamic) pressure of a gas at equilibrium can alternately be estimated from an equation of state, which involves only bulk properties of the gas, like temperature and density, and the result is identical to the mechanical pressure. However, this is no longer necessarily true for so-called *active* fluids, which are made of particles capable of autonomous motion, like self-propulsion [1], and are permanently out of thermodynamic equilibrium. Indeed, for such active fluids, pressure [2-17], but also stress [18] and chemical potential [19], may lose part (or all) of the properties they display at equilibrium and become ill-defined notions. Interestingly, the loss of equilibrium properties may depend on rather subtle details of the system. For example, active Brownian spheres still obey an equation of state when confined between torque-free walls [7], but the equation of state no longer exists when the wall is able to exert a torque on the spheres [8].

While most of the results quoted above were obtained in the limit of overdamped dynamics, we recently described and analyzed the dynamics of a gas of underdamped active dumbbells [20], which are characterized by a finite mass, a given self-propulsion force (instead of a given self-propulsion velocity) and move in a medium with finite but relatively small damping coefficient $\gamma$. The goal was to understand how the unusual properties of pressure reported in [8] emerge progressively from the characteristics of individual trajectories upon increase of the damping coefficient $\gamma$. For this purpose, a mobile wall was placed inside a 2-dimensional chamber containing an equal number of active dumbbells on each side of the wall. The repulsion force constants on both sides of the mobile wall were set to very different values and the mean position of the wall was computed for increasing values of $\gamma$ and the density of active dumbbells. The most striking result of this study was probably the observation that the displacement of the mean position of the wall is not monotonous upon increase of $\gamma$, especially at low dumbbell density, where a clear displacement first to the right and then to the left is observed [20]. The origin of this non-monotony was traced back to the existence of two different mechanisms, which are both able to disrupt the balance between the mean forces exerted by the active dumbbells on both sides of the mobile wall but tend to displace the mean position of the wall towards opposite directions.

The aim of the present paper is to extend and complete the results presented in [20] by investigating how the gas of underdamped active dumbbells recovers its thermodynamic equilibrium properties upon addition of Brownian noise. Since an ensemble of Brownian particles enclosed in an isolated chamber is at equilibrium from the thermodynamic point of view, it is expected that increasing the intensity of the Brownian noise applied to the active dumbbells masks progressively the influence of self-propulsion and enables the system to recover its equilibrium properties. Concentrating on pressure, the most naïve expectation is consequently that, upon increase of the intensity of Brownian noise, the mean position of the mobile wall returns back progressively to the center of the chamber, because the mean forces exerted by the dumbbells on both sides of the mobile wall tend to equalize. The principal result of the present paper is the demonstration that the actual dynamics of the system of underdamped active dumbbells with Brownian noise can actually be substantially more complex than this naïve prediction. For example, it will be shown that, within a certain interval of values of $\gamma$, the mean position of the mobile wall first strays further away from the center of the chamber before heading up to it upon increase of the intensity of the Brownian noise. As discussed below, detailed analysis of the dynamics of the model reveals that this complexity stems from the fact that there are actually three different mechanisms (and not only the two ones reported in [20]), which are capable of disrupting he balance between the mean forces exerted by the dumbbells on both sides of the mobile wall, and that these three mechanisms have different robustness against Brownian noise. For each value of $\gamma$ and the particle density, the relative importance of each mechanism fluctuates with increasing noise, which leads potentially to complex kinematics of the wall, because the mechanisms tend to displace the wall in different directions.

All in all, this work demonstrates that driving an autonomous system towards thermodynamic equilibrium may be a quite subtle process and that it cannot be taken for granted that equilibrium properties are recovered as a monotonous function of the driving parameter. This example may be of conceptual interest for the development of a thermodynamic theory of active matter.

The remainder of this paper is organized as follows. The model is described in Sect. II and the results of simulations performed therewith are described and interpreted in Sect. III. We finally discuss a couple of important issues and conclude in Sect. IV.

## II. Description of the model

### A. Model with two confinement chambers.

Except for the introduction of Brownian noise and the associated random forces, the model investigated in this work is similar to the one in [20] and is schematized in Fig. 1. It consists of $N$ identical self-propelled dumbbells [5,19,21-26] moving in a 2-dimensional space and enclosed between fixed walls with gross size $2L_x \times 2L_y$. A mobile wall of thickness $2e$ separates this area into two non-communicating chambers. The mobile wall can move along the $x$ axis while remaining parallel to the $y$ axis, the position of its median line being characterized by its abscissa $x_w$. This piston geometry has already been used to investigate the properties of other non-equilibrium systems, like Active Brownian Particles [8] and granular gases, either vibrated ones [27] or non-vibrated ones [28]. Corners between any two walls have the shape of a quarter of a circle of radius $r$, in order to avoid the accumulation of particles that occurs in square corners [6,29]. An equal number $N/2$ of dumbbells are enclosed in each chamber, each dumbbell $j$ being composed of two particles with respective positions $\mathbf{R}_{2j-1}$ and $\mathbf{R}_{2j}$ ($j$=1,2,…$N$) connected by a harmonic spring and separated at equilibrium by a distance $a$. Each particle experiences an active force directed along the axis of the dumbbell $\mathbf{R}_{2j} - \mathbf{R}_{2j-1}$. Moreover, each particle experiences a random force with components extracted from a Gaussian distribution. In contrast with the Active Brownian Spheres and Run-and-Tumble Particles models [30-37], where noise affects only the orientation of the velocity vector of the spheres, the Brownian force affects here all (*i.e.* translation, rotation and vibration) degrees of freedom of the active dumbbells. Besides the active and random forces, each particle also interacts with the fixed and mobile walls and with particles that do not belong to the same dumbbell through softcore potentials that vanish beyond a certain threshold and increase quadratically below the threshold.

The potential energy $V$ of the system (not including the active and random forces) is written as the sum of three terms

$$V = V_s + V_{ev} + V_w ,\tag{II-1}$$

where $V_s$ describes the internal (stretching) energy of the dumbbells, $V_{ev}$ the softcore repulsion between neighboring particles that do not belong to the same dumbbell, and $V_w$ the confining potential exerted by the walls on particles that tend to escape from the chambers. These three terms are expressed in the form

$$V_s = \frac{h}{2} \sum_{j=1}^{N} (\|\mathbf{R}_{2j-1} - \mathbf{R}_{2j}\| - a)^2$$

$$V_{ev} = \frac{h}{2} \sum_{k=1}^{2N-2} \sum_{\substack{m=k+1\,(k\,\text{even}) \\ k+2\,(k\,\text{odd})}}^{2N} H(2a - \|\mathbf{R}_k - \mathbf{R}_m\|) \times (2a - \|\mathbf{R}_k - \mathbf{R}_m\|)^2 \quad \text{(II-2)}$$

$$V_w = \frac{h_L}{2} \sum_{k \in S_L} \|\mathbf{R}_k - \mathbf{p}(\mathbf{R}_k)\|^2 + \frac{h_R}{2} \sum_{k \in S_R} \|\mathbf{R}_k - \mathbf{p}(\mathbf{R}_k)\|^2 + \frac{h}{2} \sum_{k \in S_F} \|\mathbf{R}_k - \mathbf{p}(\mathbf{R}_k)\|^2,$$

where $H(r)$ is the Heaviside step function, which insures that particles that do not belong to the same dumbbell repel each other only for separations smaller than $2a$. In the expression of $V_w$, $\mathbf{p}(\mathbf{R}_k)$ denotes the orthogonal projection of the vector coordinate $\mathbf{R}_k$ of a particle that has penetrated inside a wall on the surface of this wall (see Fig. 1), and $\|\mathbf{R}_k - \mathbf{p}(\mathbf{R}_k)\|$ represents the penetration depth of this particle inside the wall. $S_L$, $S_R$, and $S_F$ furthermore denote the sets of particles that at the considered time $t$ have penetrated inside the mobile wall coming from the left (L) and right (R) compartments and the set of particles that have penetrated inside fixed (F) walls, respectively. For the sake of simplicity, the dumbbell harmonic spring, softcore repulsive potential, and fixed wall repulsive potential share the same force constant $h$.

**B. Equations of motion.**

The equations of motion of the system are written in the form

$$m \frac{d^2 \mathbf{R}_k}{dt^2} = \mathbf{F}_k + m\gamma (v_0 \mathbf{n}_{j(k)} - \frac{d\mathbf{R}_k}{dt}) + \boldsymbol{\eta}(t)$$

$$m_w \frac{d^2 x_w}{dt^2} = F_w - m_w \gamma \frac{dx_w}{dt}, \quad \text{(II-3)}$$

($k=1,2,\ldots,2N$), where $m$ denotes the mass of the particles and $m_w$ the mass of the mobile wall, $\mathbf{F}_k$ is the force felt by particle $k$ resulting from the potential function $V$, $\gamma$ is the damping coefficient of the medium, $\mathbf{n}_j = (\mathbf{R}_{2j} - \mathbf{R}_{2j-1})/\|\mathbf{R}_{2j} - \mathbf{R}_{2j-1}\|$ the unit vector pointing from the tail to the head of dumbbell $j$, $j(k)$ denotes the integer part of $(k+1)/2$, $m\gamma v_0 \mathbf{n}_{j(k)}$ is the self-propulsion force of particle $k$, and $\mathbf{\eta}(t)$ is the noise term force vector, which components have correlation functions that satisfy

$$\langle \eta_i(t)\eta_j(t') \rangle = m^2 \gamma v_B^2 \delta_{i,j} \delta(t-t') . \tag{II-4}$$

In Eq. (II-4), $v_B^2$ characterizes the intensity of Brownian noise and can be related to a temperature $T$ through $k_B T = m v_B^2 / 2$, with $k_B$ the Boltzmann constant. For vanishing self-propulsion force ($v_0 = 0$), Brownian energy distributes equally between the various degrees of freedom of the dumbbells, with mean energies $mv_B^2/2$, $mv_B^2/4$, and $mv_B^2/4$, for translation, rotation, and vibration, respectively, whatever the value of the damping coefficient $\gamma$. In contrast, the velocity vector of isolated dumbbells ($\mathbf{F}_k = 0$ and $v_B = 0$) subject only to the self-propulsion force $m\gamma v_0 \mathbf{n}_{j(k)}$ tends to align along the geometric axis of the dumbbell with a characteristic time that decreases as $1/\gamma$, while its norm converges towards $v_0$ with the same characteristic time, so that isolated dumbbell travel asymptotically along straight lines at constant velocity $v_0$. The complexity of the dynamics that will be discussed below arises in great part from the conflict between Brownian noise, which provides the dumbbells with a rotational energy $mv_B^2/4$, and the self-propulsion force, which tends to align the velocity vector of the dumbbells along their geometric axis with a time constant $1/\gamma$. Collisions with the wall or between two dumbbells also contribute to the complexity of the dynamics by reorienting the velocity vector of the dumbbells, the mean free path between two dumbbell collisions scaling approximately like $2L_x L_y /(Na)$. For most simulations discussed below, the value of $v_B$ was increased at constant value of $v_0$, in order to decrease the importance of self-propulsion and drive the system towards thermodynamic equilibrium.

It is worth noting that Eq. (II-3) does not conserve momentum, as is also the case for the Active Brownian Spheres and Run-and-Tumble Particles models [30-37], and is consequently best suited to describe particles moving on a surface that acts as a momentum sink, like crawling cells [38] or colloidal rollers [39] and sliders [40]. However, such systems often have a large damping coefficient, while the damping coefficient is allowed here to be small. Moreover, Eq. (II-3) implies that the medium contributes to the forces exerted on the mobile wall only through its action on dumbbell dynamics. The wall is therefore assumed to be permeable to this medium and the pressure exerted by active dumbbells must be considered as an osmotic pressure [5,7].

For the purpose of numerical integration, the derivatives in Eq. (II-3) were discretized according to standard Verlet-type formulae and the equations of evolution subsequently recast into the form

$$\mathbf{R}_k^{(n+1)} = \frac{4}{2+\gamma \Delta t} \mathbf{R}_k^{(n)} - \frac{2-\gamma \Delta t}{2+\gamma \Delta t} \mathbf{R}_k^{(n-1)} + \frac{2 (\Delta t)^2}{m (2+\gamma \Delta t)} \mathbf{F}_k^{(n)}$$
$$+ \frac{2 v_0 (\Delta t)^2}{2+\gamma \Delta t} \gamma \, \mathbf{n}_{j(k)}^{(n)} + \frac{2 v_B (\Delta t)^{3/2}}{2+\gamma \Delta t} \gamma^{1/2} \, \boldsymbol{\xi}_k^{(n)} \tag{II-5}$$
$$x_w^{(n+1)} = \frac{4}{2+\gamma \Delta t} x_w^{(n)} - \frac{2-\gamma \Delta t}{2+\gamma \Delta t} x_w^{(n-1)} + \frac{2 (\Delta t)^2}{m_w (2+\gamma \Delta t)} F_w^{(n)},$$

where superscripts $(n-1)$, $(n)$, and $(n+1)$ indicate the time steps at which the quantity is evaluated, the $\boldsymbol{\xi}_k^{(n)}$ are random vectors with components extracted from a Gaussian distribution with zero mean and unit variance, and $\Delta t$ is the integration time step.

Simulations were performed with the same set of parameters as in [20], that is, $a = 1$, $L_x = L_y = 100$, $e = 8$, $r = 20$, $m = 0.5$, $m_w = 2$, and $h = h_L = 4$ but $h_R = 0.4$ to introduce a strong dissymmetry between the left and right sides of the mobile wall. $v_0$ was set to 2 in all the simulations discussed below, except for the few ones where self-propulsion was switched off ($v_0 = 0$) for the sake of comparison with the dynamics of a system at thermodynamic equilibrium. $\gamma$ was varied between 0 and 1 and $v_B$ between 0 and 2. Note that for $\gamma \approx 0.01$ the characteristic time for the alignment of the velocity vector of a particle along the tail-to-head axis of the dumbbell is of the same order of magnitude as the time it takes for the particle to cross the empty chamber at velocity $v_0$, while velocity alignment is about hundred times faster than crossing for $\gamma = 1$. Finally, simulations were performed with $N = 50$ or 500 dumbbells, which correspond to surface coverage values around 0.7% and 7%, respectively, when assuming that each particle is a disk of radius $a$. All simulations were performed with a time step $\Delta t = 0.002$.

### C. Simplified model with a single confinement chamber.

In [20], it proved useful to investigate the dynamics of a single active dumbbell enclosed inside a single chamber to understand the results of simulations performed with many dumbbells separated by a mobile wall. The same technique will be used below to decipher the dynamics of active dumbbells with Brownian noise. Briefly, the simplified model is obtained from the system described in Fig. 1 by keeping only the left confinement chamber and placing only one active dumbbell therein. Moreover, it is assumed that the interactions between the dumbbell and the four walls still obey Eq. (II-3) but the collisions of the dumbbell against the right wall cause the confinement chamber to move *as a whole* towards the right, while preserving its shape and dimensions. In contrast, the three other walls of the chamber experience no recoil upon collision with the dumbbell. For the sake of clarity, the force constant of the repulsive potential of the right wall is labeled $h_w$ (instead of $h_L$) for the modified system with a single chamber.

The complexity of the dynamics of active dumbbells subject to Brownian noise is easily grabbed by examining the properties of the trajectories of the dumbbell enclosed in a single confinement chamber, like for example its root mean square translational velocity $\sqrt{<v_t^2>}$, which is plotted in Fig. 2 for values of $v_B$ increasing from 0 to 2, as well as for the dumbbell without self-propulsion ($v_0 = 0$ and $v_B = 2$). For $v_B = 0$, the curve displays a hyperbolic tangent-like shape saturating at $v_0 = 2$ (see the related discussion in [20]), while for the dumbbell without self-propulsion the translational velocity is constant and equal to $v_B / \sqrt{2} = \sqrt{2}$. What is quite remarkable, is the fact that the average translational velocity of the active dumbbell subject to additional Brownian energy is always (except for very small values of $\gamma$) smaller than its velocity for $v_B = 0$. The reason is that the Brownian noise, as mentioned above, also contributes to increase the rotational and vibrational velocities of the dumbbell, which generates in turn more energy dissipation, because the rotational and vibrational velocities may be large even if the translational velocity is small, and because the damping force acts on each particle composing the dumbbell and not only its center of mass.

Finally, Fig. 3 gives a flavor of the extent to which the trajectories of the active dumbbell are modified upon addition of Brownian noise. Trajectories without Brownian noise ($v_B = 0$) are shown in the left column and trajectories with Brownian noise ($v_B > 0$) in the right column. It is observed that the trajectories of the dumbbell subject to Brownian noise retain some resemblance with their noiseless counterparts up to $v_B \approx 0.1$ (medium line), while the resemblance disappears rapidly for larger values of $v_B$. As seen in the bottom line of the figure, for values of $v_B$ as large as the self-propulsion velocity $v_0$, the trajectories of the active dumbbell subject to Brownian noise differ very profoundly from those of the noiseless dumbbell.

**III. Results and interpretation**

This section is devoted to the presentation and discussion of the main results obtained with the models described above upon variation of the intensity of the Brownian noise. The basic motivation underlying the simulations is to drive the system closer to thermodynamic equilibrium by increasing Brownian noise and track the transition through the variations of the mean position of the mobile wall, $\langle x_w \rangle$. At thermodynamic equilibrium, the mean forces exerted on both sides of the wall equilibrate and the wall remains on average at the center of the confinement chamber, except for tiny differences in the mean penetration depth of the particles on both sides of the wall. Significant departure of the mean position of the wall from $\langle x_w \rangle = 0$ is then a measure of the rupture of the balance between the forces exerted on both sides of the wall and, in some loose sense, of the distance of the system from thermodynamic equilibrium. The most natural expectation is consequently that, for a given value of $\gamma$ (corresponding to a certain value of $\langle x_w \rangle$ for the system without Brownian noise), $|\langle x_w \rangle|$ will progressively decrease down to 0 for increasing values of $v_B$. It will instead be shown that for certain values of $\gamma$ and for low enough dumbbell density, the evolution of the system is actually significantly more complex and can be understood only by deciphering the mechanisms, which lead to the rupture of the balance of the forces exerted on both sides of the mobile wall.

### A. Simulations with *N*=500 dumbbells.

Since they are simpler than those obtained at lower dumbbell density, let us first analyze the results of simulations performed with $N = 500$ dumbbells, which corresponds to a surface coverage value around 7%. Fig. 4 shows the evolution of the mean relative position of the mobile wall, $<x_w>/L_x$, as a function of the damping coefficient $\gamma$ for this system (note that we discuss in Sect. IV the alternative choice of $\gamma/v_B^2$ rather than $\gamma$ as the abscissa axis of the graphs for readers with a Brownian particles perspective). The plot in vignette (b) was obtained from a simulation with $v_0 = 0$ and $v_B = 2$, that is, for dumbbells without self-propulsion but subject to Brownian noise. It is seen that $\langle x_w \rangle \approx 0$ for all values of $\gamma$, as expected at thermodynamic equilibrium. The plots in vignette (a) were instead obtained from simulations with $v_0 = 2$ and values of $v_B$ ranging from 0 to 2, that is, for dumbbells out of thermodynamic equilibrium. The red solid line ($v_B = 0$) is actually a confirmation of the plot shown in Fig. 3 of [20] for dumbbells with self-propulsion but without Brownian noise. It is seen that the mean position of the mobile wall is displaced significantly towards the right for $\gamma < 0.27$ and towards the left for $\gamma > 0.27$. The mechanisms leading to the rupture of the balance between the mean forces exerted on both sides of the mobile wall were investigated thoroughly in [20]. It was found that non-monotony in the plot of $<x_w>/L_x$ versus $\gamma$ arises from the fact that two different mechanisms causing force unbalance dominate at different values of $\gamma$. More precisely, for small values of $\gamma$, dumbbells that acquire a large spin momentum upon collisions with the wall and hit it repeatedly contribute to increase significantly the mean force exerted on the harder side of the wall compared to the softer side. A typical trajectory where the dumbbell returns back rapidly towards the wall because of the spin momentum acquired during a first collision, thereby contributing efficiently to this *spin momentum mechanism*, is shown in Movie S1 [41], while the increase in wall collision frequency resulting from this mechanism will be illustrated more clearly in Sect. III.C. In contrast, with increasing $\gamma$, the mean force exerted on the softer side of the wall increases steadily compared to the harder side, because dumbbells are deflected more and more slowly away from the soft side and interact for longer and longer times with it. A typical trajectory, where the dumbbell penetrates deeply inside the soft wall, aligns slowly with it, and interacts for a long time instead of being deflected rapidly away, thereby contributing efficiently to this *slow deflection mechanism*, is shown in Movie S2 [41], while the increase in the mean duration of wall collisions resulting from this mechanism is illustrated further in Figs. S1 and S2 [41]. The spin momentum mechanism dominates (and the mean position of the wall is displaced towards the right) for $\gamma < 0.27$, while the slow deflection mechanism dominates (and the mean position of the wall is displaced towards the left) for $\gamma > 0.27$.

The other plots in vignette (a) of Fig. 4 show how the mean position of the mobile wall evolves with increasing value of $v_B$. The prominent feature is obviously that the curves tends to level off along the abscissa axis and that for $v_B \approx v_0$ the mobile wall remains close to the center of the confinement chamber in the whole range of values of $\gamma$, as for systems at thermodynamic equilibrium. Still, a closer examination of the plots indicates that the two extremities of the curve do not level off at the same rate. More precisely, the displacement of the mean position of the wall towards positive values of $\langle x_w \rangle$ around $\gamma \approx 0.18$ vanishes for $v_B = 0.5$, while for this value of $v_B$ the distance of the mean position of the mobile wall to the center of the chamber has been reduced by only 25% at $\gamma = 1$. The reason for this difference is that the two mechanisms at the origin of force unbalance do not have the same robustness against rotational noise. More precisely, as already described above, the spin momentum mechanism that prevails at low values of $\gamma$ involves dumbbells that acquire a large spin momentum when colliding with the wall and return back and hit it repeatedly (movie S1 [41]). It is consequently sufficient that rotational noise be strong enough to perturb significantly the trajectories of the dumbbells over the time interval between two collisions with the wall for this mechanism to become inefficient. In contrast, the slow deflection mechanism is based on the fact that dumbbells that collide with the softer side of the wall are deflected more slowly away from the wall and push it for longer times towards the left (movie S2). It is therefore mandatory that rotational noise be strong enough to perturb significantly the trajectories of the dumbbells over a time scale as short as the duration of one collision for this mechanism to become inefficient. Hence, the spin momentum mechanism no longer plays any role for $v_B \geq 0.5$ and the plots for larger values of $v_B$ in Fig. 4 reflect only the progressive weakening of the slow deflection mechanism with increasing Brownian noise.

In conclusion, the mean position of the mobile wall moves back towards the center of the confinement chamber with increasing Brownian noise, but the rate depends on the particular mechanism at play for the corresponding value of the damping coefficient $\gamma$.

### B. Simulations with $N$=50 dumbbells.

Let us now turn our attention to the results of simulations performed with only $N = 50$ dumbbells inside the confinement chambers, which corresponds to a surface coverage value around 0.7%, ten times smaller than in the previous sub-section. Fig. 5 shows the corresponding evolution of the mean relative position of the mobile wall, $<x_w>/L_x$, as a function of the damping coefficient $\gamma$ for $v_B = 0$ (red solid line), as well as nine other values of $v_B$ ranging from 0.05 to 2. The red solid line is again a confirmation of the plot shown in Fig. 3 of [20] for dumbbells with self-propulsion but without Brownian noise. Comparison of Figs. 4 and 5 indicates that, in the absence of Brownian noise, the mean displacement of the mobile wall away from the center of the confinement chamber is usually substantially larger for $N = 50$ than for $N = 500$ and that the evolution of $<x_w>/L_x$ versus $\gamma$ displays additional features for $N = 50$ compared to $N = 500$, namely several oscillations below $\gamma = 0.15$ and a broad bump between $\gamma = 0.4$ and $\gamma = 0.7$. Both the reduction of the mean displacement of the mobile wall and the cancellation of these additional features upon increase of $N$ from 50 to 500 are of course ascribable to the increase of the rate of collisions between dumbbells, which interfere destructively with the mechanisms leading to force unbalance. We will come back to this point in Sect. IV.

As will become clearer in Sect. III.C, the oscillations observed below $\gamma \approx 0.4$ in Fig. 5 reflect oscillations in the frequency of the collisions between active dumbbells and the mobile wall and are the signature of mechanical resonances between the rotation velocity of the dumbbells and the recoil speed of the mobile wall. As a consequence, for small values of the dumbbell density, the net effect of the spin momentum mechanism depends on the precise value of the damping coefficient. This mechanism can displace the mean position of the mobile wall either to the right (as for the main maximum around $\gamma \approx 0.2$) or to the left (as for the minimum around $\gamma \approx 0.1$). It can however be checked in Fig. 5 as well as in Fig. 6, where $<x_w>/L_x$ is plotted as a function of $v_B$ for selected values of $\gamma$, that the mean position of the mobile wall reaches the center of the confinement chamber for values of $v_B$ close to 0.2, whatever the sign of the initial displacement.

The evolution with increasing value of $v_B$ of the broad bump, which is observed between $\gamma = 0.4$ and $\gamma = 0.7$ for $v_B = 0$ (red solid line in Fig. 5), is more difficult to rationalize. It is indeed seen in Figs. 5 and 6 that for $\gamma \approx 0.5$ the mean position of the mobile wall first strays significantly further away from the center of the confinement chamber for values of $v_B$ up to about 0.05, before moving back towards the center of the chamber for values of $v_B$ larger than about 0.10. This evolution cannot be explained by invoking only the two mechanisms described above, because for such values of $\gamma$ the spin momentum is damped too rapidly to play any role, while the progressive weakening of the slow deflection mechanism can hardly explain the forth and back displacements of the mean position of the wall. By the way, no attempt was made in [20] to understand the origin of the bump in the plot of $<x_w>/L_x$ versus $\gamma$, while it appears here that it may be the fingerprint of an important aspect of the dynamics of the active dumbbells.

This feature points clearly towards the need for a deeper understanding of the dynamics of the active dumbbells. As discussed below, this was achieved through the investigation of the dynamics of the simplified model described in Sect. II.C, which consists of a single active dumbbell enclosed inside a single confinement chamber.

### C. Simulations with the simplified model with a single confinement chamber.

Simulations with the simplified model were performed with two different values of $h_w$, namely 4.0 and 0.4, which correspond, respectively, to the values of $h_L$ and $h_R$ for the complete model. The average displacement per unit time of the chamber towards the right was computed for each set of values of $h_w$, $\gamma$, and $v_B$, and the value obtained for $h_w = 0.4$ was subtracted from the value obtained for $h_w = 4.0$, thus yielding a differential displacement per unit time labeled $\Delta v_w$. This quantity is plotted in Fig. 7 as a function of $\gamma$ for values of $v_B$ ranging from 0 to 2. The very thin peaks, which appear in the plot obtained for the noiseless system ($v_B = 0$, the solid red line in vignette (a) of Fig. 7), were shown in [20] to be the fingerprint of pseudoperiodic trajectories. These pseudoperiodic trajectories are destroyed (and the peaks disappear from the plots) for the weakest Brownian noise used in the present study ($v_B = 0.005$, not shown in Fig. 7) and will not be discussed any further below.

Beside these thin peaks, Fig. 7 displays features that compare readily to those of Fig. 5. More precisely, the broad maximum and the narrower oscillations observed at low damping coefficient ($\gamma \leq 0.4$), which are due to the spin momentum mechanism, level off for $v_B \approx 0.2$, while the negative values of $\Delta v_w$ observed at larger damping coefficients, which are due to the slow deflection mechanism, cancel for values of $v_B$ as large as the self-propulsion velocity $v_0 = 2$. The top and bottom lines in Fig. 3 illustrate the extent to which the trajectories of the active dumbbell need be perturbed by the Brownian noise for the mean forces exerted on walls with very different repulsion force constants to equilibrate. Quite interestingly, the plot of $\Delta v_w$ versus $\gamma$ displays an additional local maximum around $\gamma \approx 0.55$, which levels off for values of $v_B$ as small as $v_B \approx 0.1$ and is obviously the counterpart of the broad bump observed in Fig. 5. It is the mechanism leading to this maximum that we must strive to understand.

A first indication is provided by the plots of the scaled collision frequency $f / \sqrt{\langle v_t^2 \rangle}$ as a function of $\gamma$ for values of $v_B$ ranging from 0 to 2, which are shown in Figs. 8 and 9 for $h_w = 0.4$ and $h_w = 4.0$, respectively. $f$ is the number of times the dumbbell hits the right wall per unit time, and $\sqrt{\langle v_t^2 \rangle}$ the root mean square translational velocity of the dumbbell (see Fig. 2). Figs. 8 and 9 show the evolution with $\gamma$ of $f / \sqrt{\langle v_t^2 \rangle}$ instead of $f$, because the collision frequency increases almost linearly with the translational velocity of the dumbbell and the plots are clearer after removal of this dependence. In both figures, the broad maximum centered around $\gamma \approx 0.2$ on the curve for $v_B = 0$ reflects the increase in collision frequency due to the spin momentum mechanism. Consequently, the oscillations observed on the left aisle of the broad peak are the hallmarks of mechanical resonances between the rotation velocity of the dumbbell and the recoil speed of the confinement chamber, as mentioned in Sect. III.B. However, the most salient feature of these figures is the fact that addition of Brownian noise induces a gradual decrease in the collision frequency for $\gamma \leq 0.3$, which reflects the weakening of the spin momentum mechanism by rotational noise, as discussed above, while for $\gamma > 0.3$ a significant *increase* in the collision frequency is instead observed upon increase of the Brownian noise up to $v_B \approx 0.1$. This effect is particularly marked for $h_w = 0.4$, the collision frequency being multiplied by a factor close to 2 for $\gamma \approx 0.7$ and $v_B \approx 0.1$ compared to $v_B = 0$. Examination of the trajectories of the dumbbell for such values of $\gamma$ and $v_B = 0$ suggests that the dumbbell spend most time travelling parallel to the $y$ axis, thereby experiencing relatively few collisions with the right wall of the chamber, while addition of a small amount of Brownian noise is sufficient to reorient the translational velocity almost randomly, thereby increasing the frequency of the collisions with this wall. This hypothesis can be checked quantitatively by plotting the probability density $p(\theta)$ for the velocity vector of the center of mass of the dumbbell to be oriented with an angle $\theta$ with respect to the $x$ axis for increasing values of $v_B$. $p(\theta)$ is shown in Fig. 10 for $\gamma = 0.70$ and $h_w = 0.4$ (top plot) or $h_w = 4.0$ (bottom plot). It is seen that the probability density is indeed strongly peaked around $\theta \approx \mp \pi/2$ for $v_B = 0$, especially for $h_w = 0.4$. Increasing $v_B$ up to 0.05 (for $h_w = 4.0$) or 0.10 (for $h_w = 0.4$) is, however, sufficient to let the distribution become almost flat, meaning that the dumbbell has no longer any preferential orientation.

### D. A third mechanism for the rupture of force balance.

These simulations therefore pinpoint the existence of a third mechanism leading to the rupture of the balance of the mean forces exerted on both sides of the mobile wall when the active dumbbell is not subject to Brownian noise. Two mechanisms were identified in [20], namely the spin momentum mechanism, which is efficient at low values of the damping coefficient, and the slow deflection mechanism, which efficiency increases with the damping coefficient, but this third mechanism, which is efficient at intermediate values of $\gamma$, was missed. This mechanism relies on the preferential alignment of the trajectories parallel to the mobile wall (but not in contact with the wall) in the absence of Brownian noise, due probably both to the geometry of the confinement chambers (in particular the rounded corners) and the recoil of the mobile wall when hit by a dumbbell. $p(\theta)$ is more strongly peaked around $\theta \approx \mp\pi/2$ for $h_w = 0.4$ than for $h_w = 4.0$, so that the number of collisions and the mean force exerted on the wall are reduced more significantly on the softer side of the wall than on its harder side. Quite interestingly, this *parallel alignment* mechanism tends to decrease the frequency of collisions on both sides of the mobile wall, while the spin momentum mechanism tends to increase the frequency of collisions on both sides of the wall. Moreover, the parallel alignment mechanism displaces the mean position of the mobile wall towards the right (positive values of $<x_w>$), while the slow deflection mechanism displaces it towards the left (negative values of $<x_w>$), and the precise effect of the spin momentum mechanism depends on the exact value of the damping coefficient (for $N = 50$, $<x_w>$ is negative for $\gamma = 0.13$ and positive for $\gamma = 0.21$, see Figs. 5 and 6). Like the spin momentum mechanism, the parallel alignment mechanism is very sensitive to Brownian noise, because it suffices that rotational noise perturbs the trajectory significantly over the time interval it takes for the dumbbell to cross the chamber for the mechanism to become ineffective, which is the case for values of $v_B$ as small as $v_B \approx 0.1$.

Practically, for the system with $N = 50$ active dumbbells without Brownian noise and for a damping coefficient around $\gamma \approx 0.50$, the balance of the mean forces exerted on both sides of the mobile wall is disrupted by both the parallel alignment mechanism, which tends to displace the mean position of the wall towards the right, and the slow deflection mechanism, which tends to displace the mean position of the wall towards the left. The effect of the second mechanism being larger than the effect of the first one, $<x_w>$ is negative for $v_B = 0$. As discussed above, Brownian noise with weak intensity ($v_B \approx 0.1$) is sufficient to spoil the parallel alignment mechanism, while it has little effect on the slow deflection mechanism. As a result, the mean position of the wall moves to even more negative values of $<x_w>$ and strays further away from the center of the chamber. However, for $v_B > 0.1$ there consequently remains only one mechanism causing force unbalance, namely the slow deflection mechanism, so that the mean position of the mobile wall subsequently moves steadily back towards the center of the chamber upon further increase of $v_B$.

**IV. Discussion and conclusion**

In this paper, we have studied numerically the properties of a gas of underdamped self-propelled dumbbells subject to Brownian noise of increasing intensity, in order to understand how the notion of mechanical pressure is gradually recovered as the system approaches thermodynamic equilibrium. For this purpose, we considered an equal number of active dumbbells enclosed in the two chambers of a two-dimensional container separated by a mobile asymmetric wall. Since the dumbbells are capable of self-propulsion and the walls exert a torque on them during collisions, the system is not at thermodynamic equilibrium, the balance of mean forces exerted on both sides of the asymmetric wall is broken, and mechanical pressure cannot be defined properly [8]. As a consequence, the mean position of the mobile wall does not coincide with the center of the container and depends on several factors, including the asymmetry of the wall, the damping coefficient, and the density of dumbbells. These out-of-equilibrium properties have been investigated in detail in [20]. In the present work, we have extended and completed this first study by subjecting the active dumbbells to increasing Brownian noise, thereby driving the system closer and closer to a gas of Brownian particles, which is at equilibrium from the thermodynamic point of view, and analyzing the response of the system to such increasing noise. The simulations confirm that the mean forces exerted on both sides of the asymmetric wall equilibrate progressively and that the notion of mechanical pressure is recovered for sufficiently strong noise. The simulations also highlight more subtle properties of the system. First, Brownian noise of moderate intensity ($v_B \ll v_0$) is sufficient to let the mean forces equilibrate for small values of the damping coefficient (up to $\gamma \approx 0.2$), while much stronger noise ($v_B \approx v_0$) is required for larger values of the damping coefficient. Moreover, the displacement of the mean position of the mobile wall towards the center of the container upon increase of the noise intensity may not be monotonous but subject instead to changes of direction. Examination of the dynamics of the system has revealed that both phenomena relate to the mechanisms leading to the rupture of force balance and, more precisely, to the fact that there actually exist several mechanisms, which tend to displace the mean position of the wall towards different directions and display different robustness against an increase of the intensity of Brownian noise.

Several points are worth commenting before concluding this work.

First, it is important to stress that the results obtained here depend only marginally on the value of the mass of the mobile wall. A doubt might arise, because standard results are usually obtained under the assumption that walls are heavy and move slowly (if at all) compared to gas particles, while the mobile wall is assumed here to have a mass comparable to that of the dumbbells. To check this point, a series of simulations were performed with a wall mass ten times larger ($m_w = 20$ instead of $m_w = 2$). While results differ slightly from those obtained with a lighter wall, the general trends and conclusions remain valid. As an illustration, the plot of $<x_w>/L_x$ as a function of $\gamma$ for $N = 500$ active dumbbells is shown in Fig. 11 for $v_0 = 2$, $v_B = 0$, and $m_w = 2$ or $m_w = 20$. It is observed that the displacement of the mean position of the wall towards positive values of $<x_w>$ for small values of $\gamma$ and towards negative values of $<x_w>$ for larger values of $\gamma$ is preserved, in spite of the fact that the amplitudes of the displacements differ somewhat, especially at small values of $\gamma$. Similarly, the results described above depend only slightly on the internal degree of freedom of the dumbbells, that is the distance between the two particles. The force constant of the spring connecting these particles ($h = 4$ in the expression of $V_s$ in Eq. (II-2)) is indeed large enough for the dumbbells to be only slightly compressed during most relevant events. Moreover, the period of free vibration $2\pi(m/2h)^{1/2} = \pi/2$ corresponds approximately to the translation of the dumbbells over their own length and is small compared to most time scales of the system. The dumbbells are therefore expected to behave approximately like rigid rods with small aspect ratio, although the description in terms of particles and springs was used to avoid having to deal with the more involved equations for rigid bodies. As a check, the evolutions with $\gamma$ of the mean position of the mobile wall for $N = 50$ active dumbbells are compared in Fig. 12 for spring force constants $h = 4$ (as for all other simulations of this work) and $h = 40$. It is seen that the mean position of the wall depends only slightly on $h$ at low values of $\gamma$, where the spin momentum mechanism dominates, as well as at large values of $\gamma$, where the slow deflection mechanism dominates. In contrast, significant differences are observed at intermediate values of $\gamma$, where the parallel alignment mechanism dominates, which confirms (see Sect. III.D) that this mechanism is indeed rather sensitive to very small perturbations of the system.

Regarding the role of the geometry of the confinement chamber, it is worth emphasizing that both the spin momentum mechanism and the slow deflection mechanism take place in the vicinity of the mobile wall (see Movies S1 and S2 [41]), so that there is no reason why they should be significantly affected by any reasonable change in the geometry of the confinement chamber (this holds for the mechanical resonances observed at small values of $\gamma$). In contrast, changes in the geometry may affect the frequency of occurrence of collisions contributing to the spin momentum mechanism or the slow deflection mechanism, for example by modifying the statistical weight of the parallel alignment mechanism discussed in Sect. III.D or the distribution of incidence angles. Additional simulations with a different geometry ($L_x = 50$ and $L_y = 200$, instead of $L_x = L_y = 100$) were performed to assess this point. The results shown in Fig. 13 indicate that the mean displacements of the mobile wall follow closely those obtained with the original geometry, albeit with a somewhat reduced amplitude for all values of $\gamma$. This probably reflects the fact that dumbbells travel more parallel to the mobile wall, which in turn decreases the efficiency of the two mechanisms leading to the rupture the balance of mean forces.

Let us now comment succinctly on the influence of the dumbbell density. Simulations with the complete model were performed with $N = 50$ or 500 dumbbells enclosed inside the confinement chamber, which corresponds to surface coverage values around 0.7% and 7%, respectively, when assuming that each particle is a disk of radius $a$. These simulations therefore pertain clearly to the dilute regime. More important, however, is the fact already pointed out in Sect. II.B that the mean free path between two dumbbell collisions is approximately equal to $2L_xL_y/(Na)$, that is $4L_x$ for $N = 50$ and $0.4L_x$ for $N = 500$. This agrees with the observation that for $N = 50$ most dumbbells cross the confinement chambers without experiencing collisions with other dumbbells, while they are usually involved in several collisions for $N = 500$. The results obtained with $N = 50$ dumbbells are therefore expected to be very close to the infinite dilution limit for this precise geometry, while collisions among dumbbells do affect significantly the results for $N = 500$. Collisions and Brownian noise share some common points, in the sense that both of them perturb the trajectories of the dumbbells in a random-like fashion. However, Brownian noise affects the trajectories continuously, while collisions act over the narrow time window during which colliding particles interact. As a consequence, these two perturbations do not weaken the mechanisms leading to the rupture of force balance in a similar way. For example, comparison of Figs. 4 and 5 indicates that introducing $N = 500$ dumbbells in the confinement chamber has the same effect as adding Brownian noise with $v_B \approx 0.07$ at $\gamma \approx 0.2$, where the spin momentum mechanism dominates, but with $v_B \approx 0.40$ at $\gamma = 1.0$, where the slow deflection mechanism dominates. Similarly, comparison of Fig. 3 of [20] and Fig. 5 indicates that introducing $N = 5000$ dumbbells in the confinement chamber, which corresponds to a surface coverage around 70%, has the same effect on the slow deflection mechanism as adding Brownian noise with $v_B \approx 1.0$.

Last but not least, let us mention that the results presented above may be exploited and interpreted in a slightly different manner when considering that the model actually consists of Brownian dumbbells perturbed by self-propulsion, instead of self-propelled dumbbells perturbed by Brownian noise, which is the implicit point of view that was assumed throughout the paper. From the point of view of Brownian particles, the strength of the perturbation induced by self-propulsion can be estimated from the Péclet number, which is the ratio of the transport rate due to self-propulsion to the rate of diffusion $D = (mv_B^2)/(2m\gamma)$, that is $\text{Pe} = av_0/D = 2\gamma a v_0 / v_B^2$. Plotting the results of the simulations as a function of $\gamma$ for increasing values of $v_B$, as was done in most figures, therefore looks like performing convoluted cuts through the parameter space, since perturbation is expected to increase like $\gamma/v_B^2$. From this perspective, it consequently appears more natural to plot the results as a function of $\gamma/v_B^2$ rather than $\gamma$. To check this viewpoint, the evolution of the mean relative position of the mobile wall for $N = 50$ dumbbells is shown in Fig. 14 as a function of $\gamma/v_B^2$, while it was plotted as a function of $\gamma$ in Fig. 5. If the response of the system were controlled by the Péclet number, then all curves would collapse onto a single master curve, while approximate collapse is observed in Fig. 14 only for values of $v_B$ larger than 1, that is in a regime where the only mechanism leading to the rupture of force balance is the slow deflection mechanism. Conversely, the spin momentum mechanism is effective only for very large values of Pe.

In conclusion, this work provides a clear illustration of the fact that driving an autonomous system towards (or away from) thermodynamic equilibrium may not be a straightforward process, but may instead proceed through variations of the relative weights of several conflicting mechanisms. It should be stressed that, in the context of the model developed in [20] and the present paper, interesting (complex) behavior is observed essentially for small values or the damping coefficient, because up to three different mechanisms leading to the rupture of force balance may superpose for small values of $\gamma$, while a single one is active for $\gamma \approx 1$. This decrease in the number of mechanisms with increasing damping coefficient is in turn directly related to the drastic simplification of the trajectories of the noiseless dumbbells, which evolve from intricate and sensitive volutes at small $\gamma$ to essentially straight lines at large $\gamma$, and is consequently likely to be quite general (model-independent). As already mentioned, the model for dry active matter discussed here is however best suited to describe particles moving on a surface that acts as a momentum sink, like crawling cells or colloidal rollers and sliders, but these systems usually have a large damping coefficient. While the results described in the present paper are interesting from the conceptual point of view, it may consequently prove difficult to detect realizations thereof in real systems. On the other hand, wet active matter, like bacteria swimming in the bulk or suspensions of catalytic colloidal rods, may have much smaller damping coefficients. Although the equations of motion are different (and more demanding from the numerical point of view), it is probable that the superposition of mechanisms leading to the rupture of force balance described here in the context of dry active matter is also effective for these wet active systems. This may be a point worth remembering, or eventually even checking, when studying such systems.

**FIGURE CAPTIONS**

**Figure 1** : **(a)** Schematic diagram of a dumbbell, showing the two particles located at positions $\mathbf{R}_{2j-1}$ (tail) and $\mathbf{R}_{2j}$ (head), the string connecting them, and the self-propulsion force applied to each particle and directed from the tail to the head of the dumbbell. **(b)** Schematic diagram of the confinement chambers. Fixed walls are shown as black solid lines and the mobile wall as red dotted lines. $x_w$ denotes the abscissa of the median line of the mobile wall. Also shown is the position $\mathbf{R}_k$ of a particle that has penetrated inside a fixed wall and its projection $\mathbf{p}(\mathbf{R}_k)$ on the surface of the wall. The repelling force exerted by the wall on this particle is proportional to $\|\mathbf{R}_k - \mathbf{p}(\mathbf{R}_k)\|$. The force constants associated with the repulsion potential on the left side of the mobile wall ($h_L$) and the right side of the mobile wall ($h_R$) are different.

**Figure 2** : Evolution, as a function of the damping coefficient $\gamma$, of the root mean square translational velocity $\sqrt{<v_t^2>}$ of a single dumbbell enclosed in a single confinement chamber for $v_0 = 2$ and 14 values of $v_B$ ranging from 0.0 to 2.0, as well as for the system without self-propulsion ($v_0 = 0$ and $v_B = 2$), see the legend. Note that the four plots for values of $v_B$ ranging from 0.0 to 0.05 nearly superpose. The plots shown here were obtained for $h_w = 4.0$, but the plots obtained for $h_w = 0.4$ are almost identical. Each plot was obtained by integrating the equations of motion for $2.0 \ 10^{12}$ time steps, with $\gamma$ increasing regularly between 0 and 1, and averaging $\sqrt{<v_t^2>}$ over $2.0 \ 10^9$ successive steps.

**Figure 3** : Representative trajectories of the active dumbbell without Brownian noise ($v_B = 0$, left column) and with Brownian noise ($v_B > 0$, right column) for the modified model with a single confinement chamber and (a) $h_w = 4.0$ and $\gamma = 0.21$, (b) $h_w = 0.4$ and $\gamma = 0.52$, and (c) $h_w = 0.4$ and $\gamma = 1.00$. The confinement chamber moves towards the right each time it is hit by the dumbbell. Represented in this figure are only its initial (in blue) and final (in red) positions. Each trajectory is integrated for 1000 time units.

**Figure 4** : Evolution, as a function of the damping coefficient $\gamma$, of the mean relative position of the mobile wall $<x_w>/L_x$ for $N = 500$ active dumbbells obeying Eq. (II-3) obtained from simulations with (a) $v_0 = 2$ and values of $v_B$ ranging from 0 to 2 (see the legend), or (b) $v_0 = 0$ and $v_B = 2$. The plot in (b) corresponds to a system at thermodynamic equilibrium and those in (a) to systems out of thermodynamic equilibrium. Each plot was computed from a single simulation integrated for $5.0\ 10^9$ steps, with $\gamma$ increasing regularly from 0 to 1 and $<x_w>$ being averaged over $5.0\ 10^7$ successive steps.

**Figure 5** : Evolution, as a function of the damping coefficient $\gamma$, of the mean relative position of the mobile wall $<x_w>/L_x$ for $N = 50$ active dumbbells obeying Eq. (II-3) with $v_0 = 2$ and 10 different values of $v_B$ ranging from 0 to 2 (see the legend). Each plot was computed from the average of 8 different simulations integrated for $10^{10}$ steps, with $\gamma$ increasing regularly from 0 to 1 and $<x_w>$ being averaged over $10^8$ successive steps. See Fig. 14 for a graph showing the same results plotted as a function of $\gamma/v_B^2$ instead of $\gamma$ and Sect. IV for a discussion of the choice of the abscissa axis.

**Figure 6** : Evolution, as a function of the mean random velocity $v_B$, of the mean relative position of the mobile wall $<x_w>/L_x$ for $N=50$ active dumbbells obeying Eq. (II-3) with $v_0=2$ and 5 different values of the damping coefficient $\gamma$ ranging from 0.13 to 1.0 (see the legend). Each plot was computed from the average of 8 different simulations integrated for $10^{10}$ steps, with $v_B$ increasing regularly from 0 to 2 and $<x_w>$ being averaged over $10^8$ successive steps.

**Figure 7** : Evolution of $\Delta v_w$ as a function of the damping coefficient $\gamma$ for $v_0=2$ and values of $v_B$ ranging (a) from 0.0 to 0.3, and (b) from 0.5 to 2.0, see the legend. Also shown as a red solid line in (b) is the plot obtained for the system without self-propulsion, that is for $v_0=0$ and $v_B=2$. These plots were obtained for the modified model with a single confinement chamber and a single active dumbbell enclosed therein. $\Delta v_w$ is the difference between the values of the mean displacement of the chamber towards the right per unit time obtained for $h_w=4.0$ and $h_w=0.4$. See the caption of Fig. 2 for computational detail.

**Figure 8** : Evolution of $f/\sqrt{\langle v_t^2 \rangle}$ as a function of the damping coefficient $\gamma$ for $h_w=0.4$, $v_0=2$ and values of $v_B$ ranging (a) from 0.0 to 0.3, and (b) from 0.5 to 2.0, see the legend. Also shown as a red solid line in (b) is the plot obtained for the system without self-propulsion, that is for $v_0=0$ and $v_B=2$. These plots were obtained for the modified model with a single confinement chamber and a single active dumbbell enclosed therein. $f$ is the number of times the dumbbell hits the right wall per unit time and $\sqrt{\langle v_t^2 \rangle}$ the root mean square translational velocity of the dumbbell (see Fig. 2). The horizontal gray dot-dashed line is just a guideline for the eyes. See the caption of Fig. 2 for computational detail.

**Figure 9** : Same as Fig. 8, but for $h_w = 4.0$ instead of $h_w = 0.4$.

**Figure 10** : Probability density $p(\theta)$ for the velocity vector of the center of mass of the dumbbell to be oriented with an angle $\theta$ with respect to the $x$ axis for $\gamma = 0.70$, values of $v_B$ in the range $0 \leq v_B \leq 0.3$, and (a) $h_w = 0.4$ or (b) $h_w = 4.0$. The plots are symmetric with respect to the $\theta = 0$ axis. They were obtained for the modified model with a single confinement chamber and a single active dumbbell enclosed therein. Each plot was computed by integrating the equations of motion for $3.0\ 10^{11}$ time steps, with $v_B$ increasing regularly between 0 and 0.3, and averaging $p(\theta)$ over $10^9$ successive steps for each bin.

**Figure 11** : Evolution, as a function of the damping coefficient $\gamma$, of the mean relative position of the mobile wall $<x_w>/L_x$ for $N = 500$ active dumbbells obeying Eq. (II-3) with $v_0 = 2$ and $v_B = 0$. The solid red line was obtained for a wall mass $m_w = 2$, as all other simulations discussed in this paper, while the dashed blue line was obtained for $m_w = 20$. Each plot was computed from the average of five simulations integrated for $5.0\ 10^9$ steps, with $\gamma$ increasing regularly from 0 to 1 and $<x_w>$ being averaged over $5.0\ 10^7$ successive steps. The two plots run almost parallel to each other for values of $\gamma$ in the range $0.35 \leq \gamma \leq 1.0$ (not shown).

**Figure 12** : Evolution, as a function of the damping coefficient $\gamma$, of the mean relative position of the mobile wall $<x_w>/L_x$ for $N = 50$ active dumbbells obeying Eq. (II-3) with $v_0 = 2$ and $v_B = 0$. The solid red line was obtained for $h = 4$ in the expression of $V_s$ in Eq.

(II-2), as all other simulations discussed in this paper, while the dashed blue line was obtained for $h = 40$. Each plot was computed from the average of 8 different simulations integrated for $10^{10}$ steps, with $\gamma$ increasing regularly from 0 to 1 and $<x_w>$ being averaged over $10^8$ successive steps.

**Figure 13** : Evolution, as a function of the damping coefficient $\gamma$, of the mean relative position of the mobile wall $<x_w>/L_x$ for $N = 50$ active dumbbells obeying Eq. (II-3) with $v_0 = 2$ and $v_B = 0$. The solid red line was obtained for $L_x = L_y = 100$, as all other simulations discussed in this paper, while the dashed blue line was obtained for $L_x = 50$ and $L_y = 200$. Each plot was computed from the average of 8 different simulations integrated for $10^{10}$ steps, with $\gamma$ increasing regularly from 0 to 1 and $<x_w>$ being averaged over $10^8$ successive steps.

**Figure 14** : Same as Fig. 5, except that $<x_w>/L_x$ is plotted as a function of $\gamma/v_B^2$ instead of $\gamma$. The horizontal axis is consequently proportional to the Péclet number $\text{Pe} = av_0/D = 2\gamma av_0/v_B^2$, that is the dimensionless swimming speed. It is natural to use $\gamma/v_B^2$ instead of $\gamma$ as the horizontal axis when considering that the model actually consists of Brownian dumbbells perturbed by self-propulsion, rather than self-propelled dumbbells perturbed by Brownian noise. For a given value of $\gamma/v_B^2$, the effects of inertia increase with decreasing value of $v_B$.

**Figure 1**

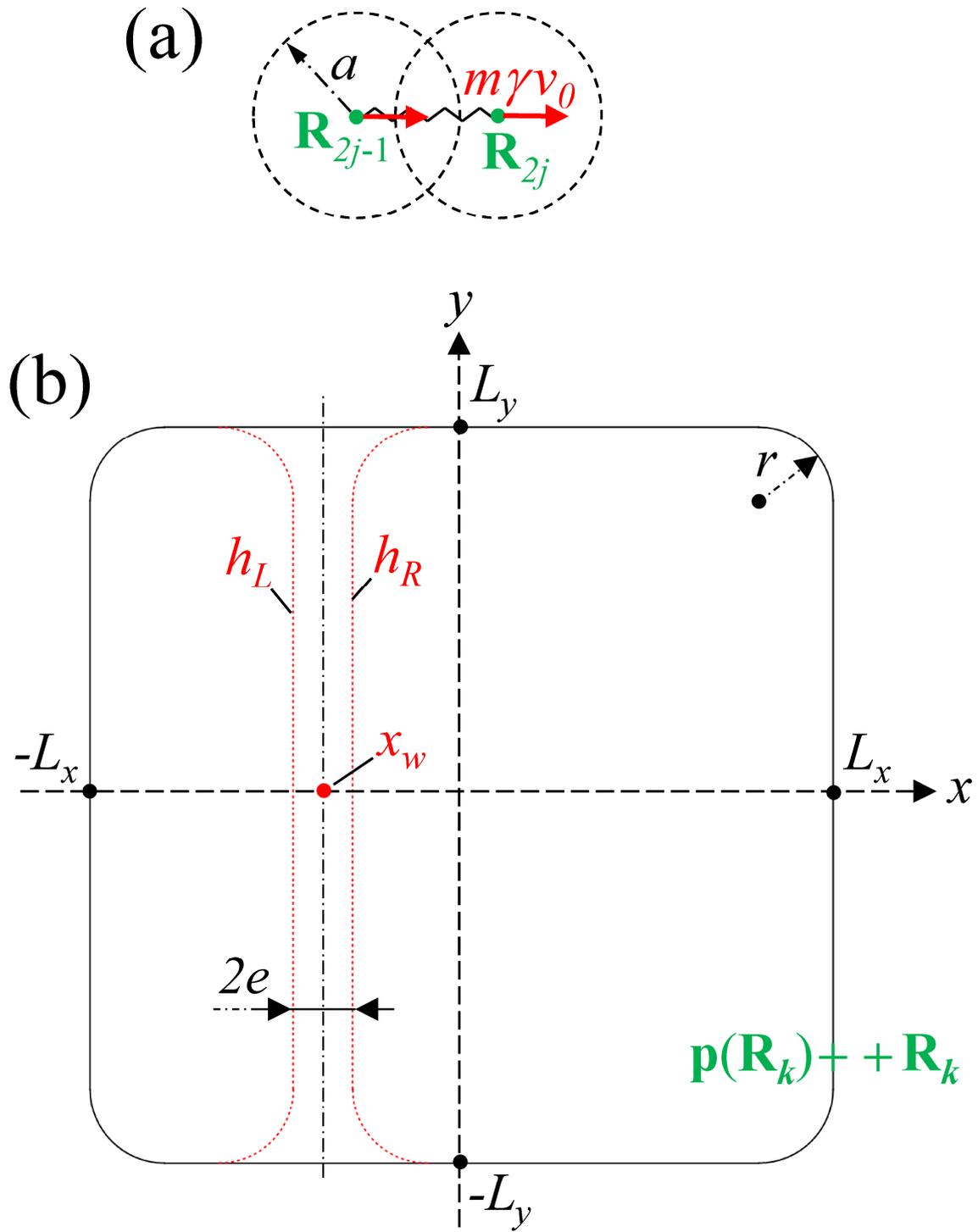

**Figure 2**

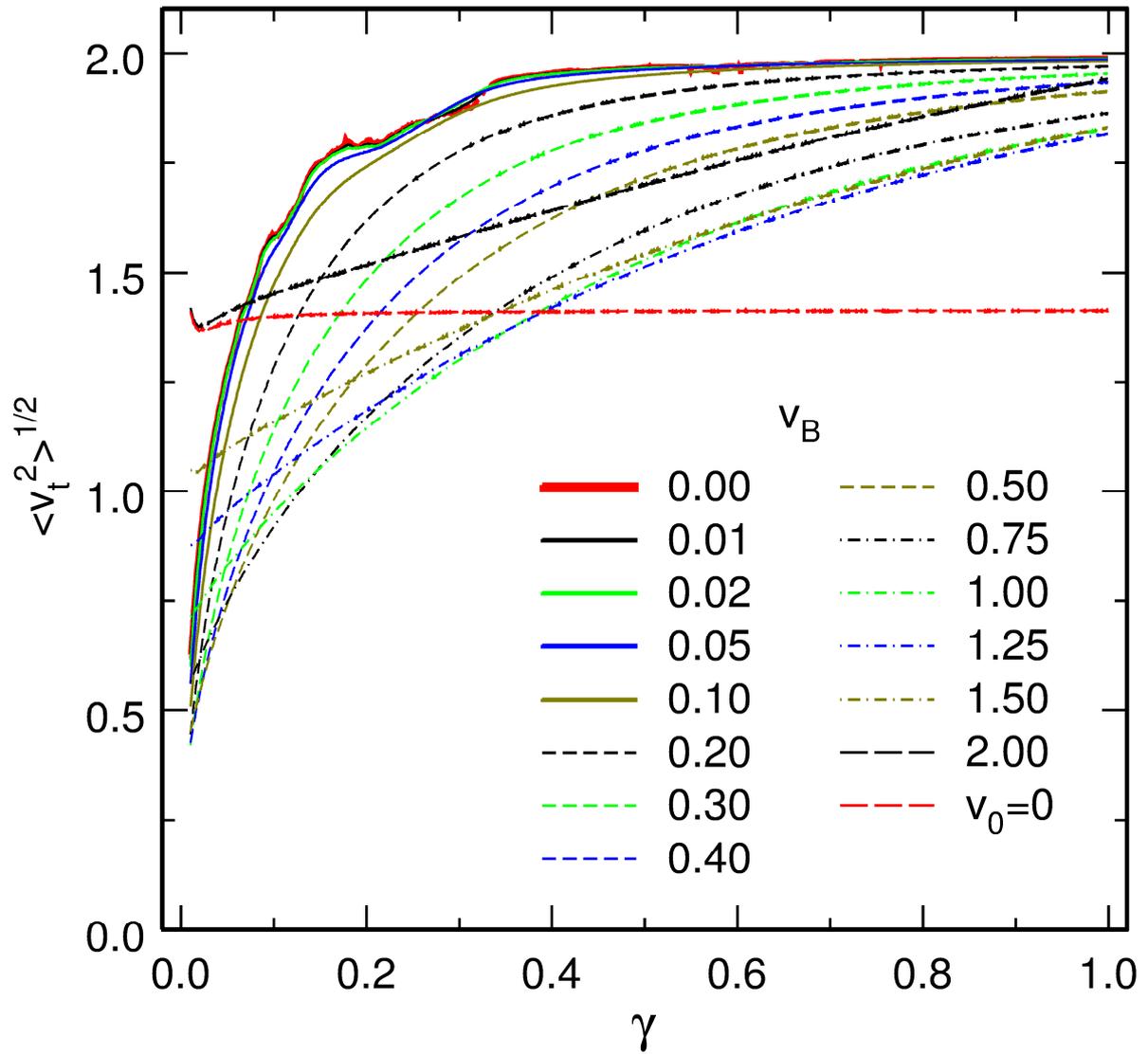

**Figure 3**

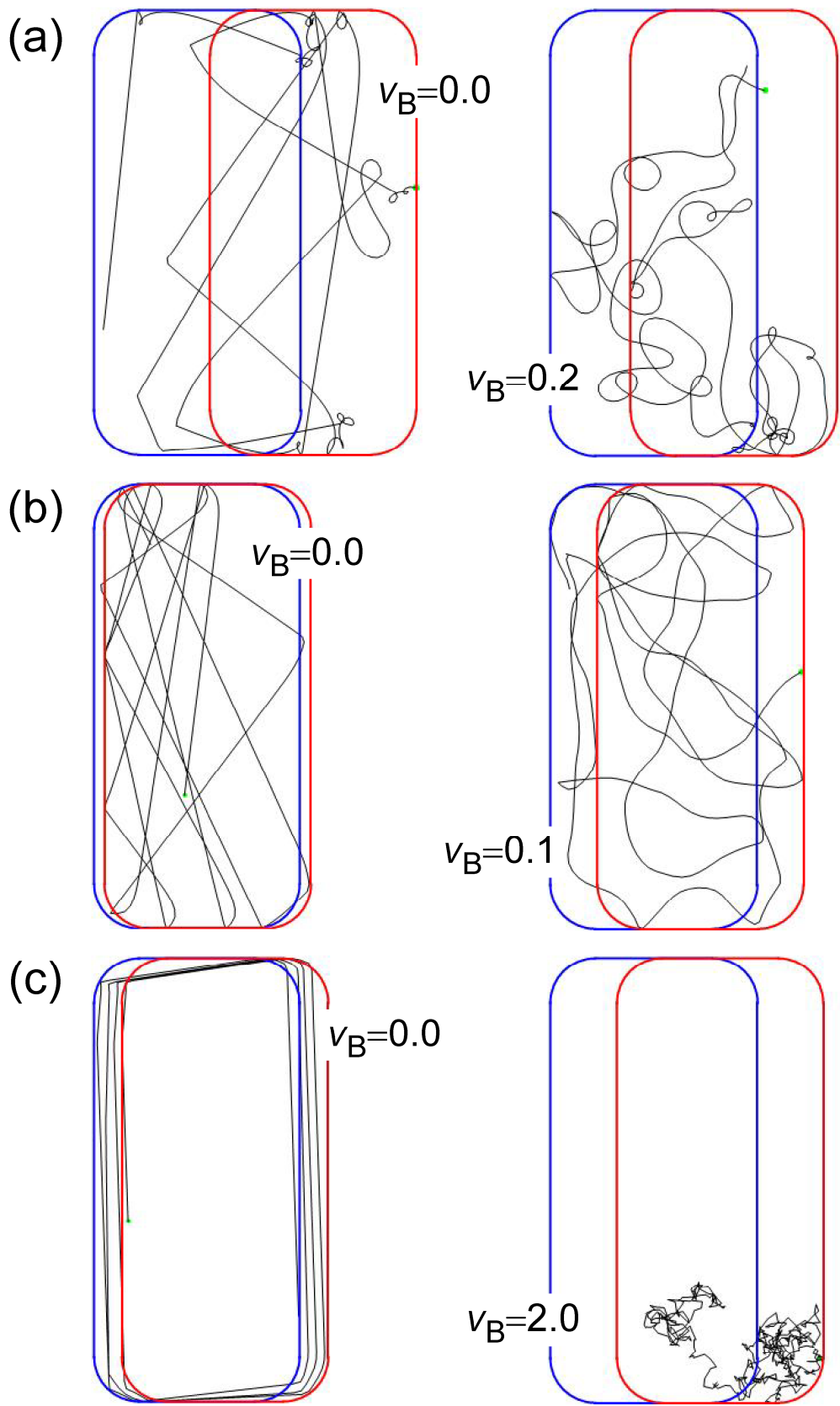

**Figure 4**

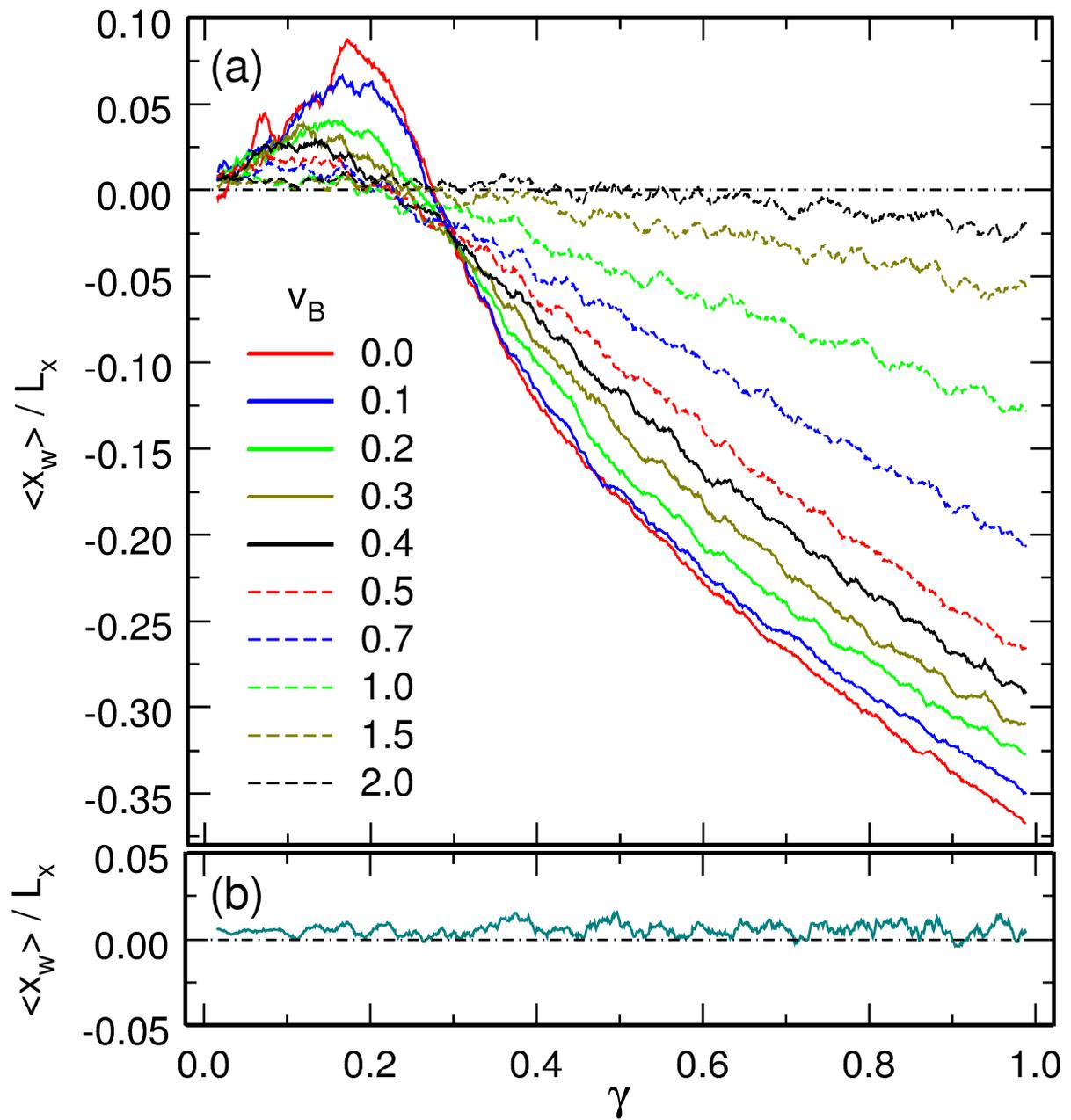

**Figure 5**

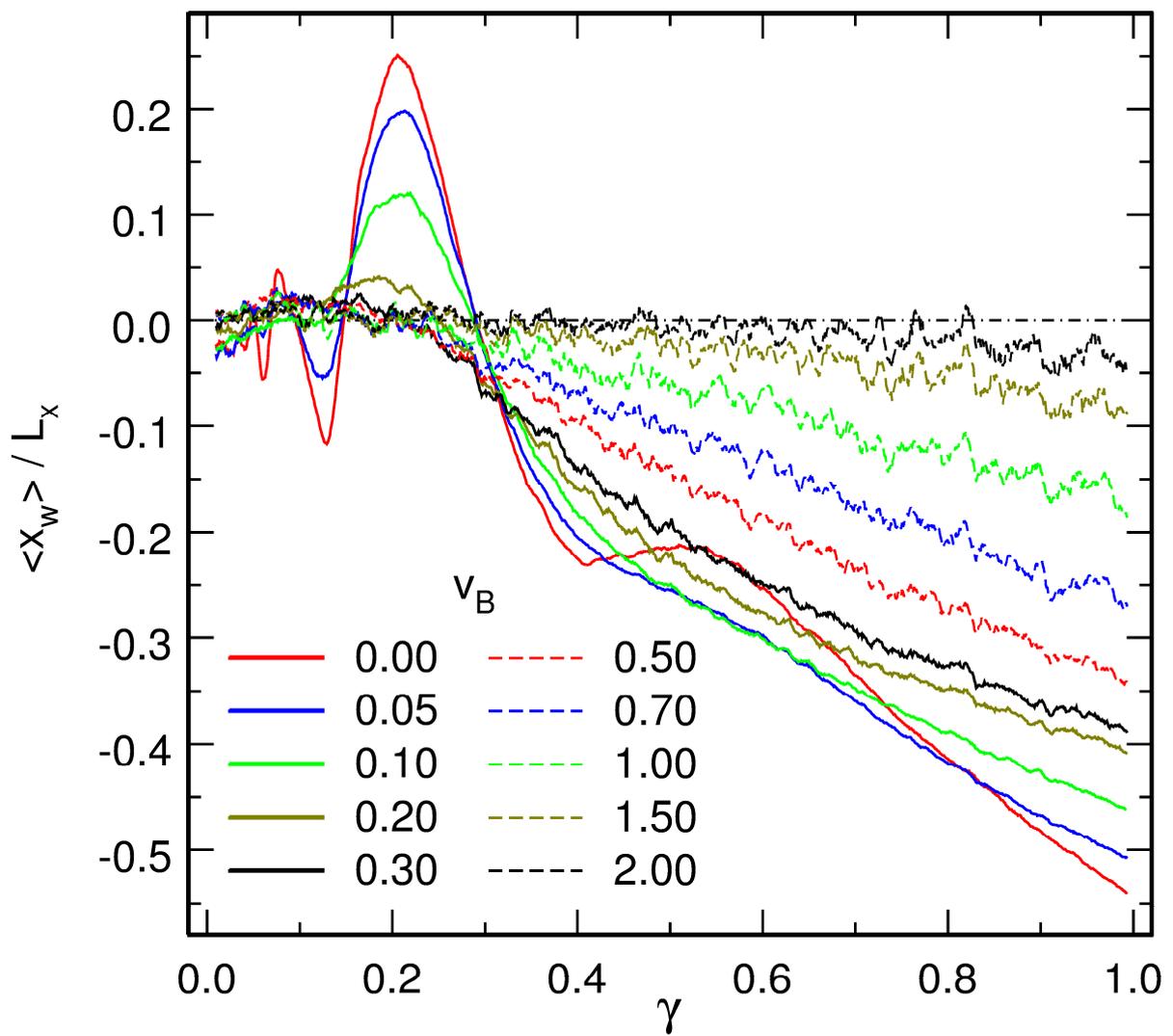

**Figure 6**

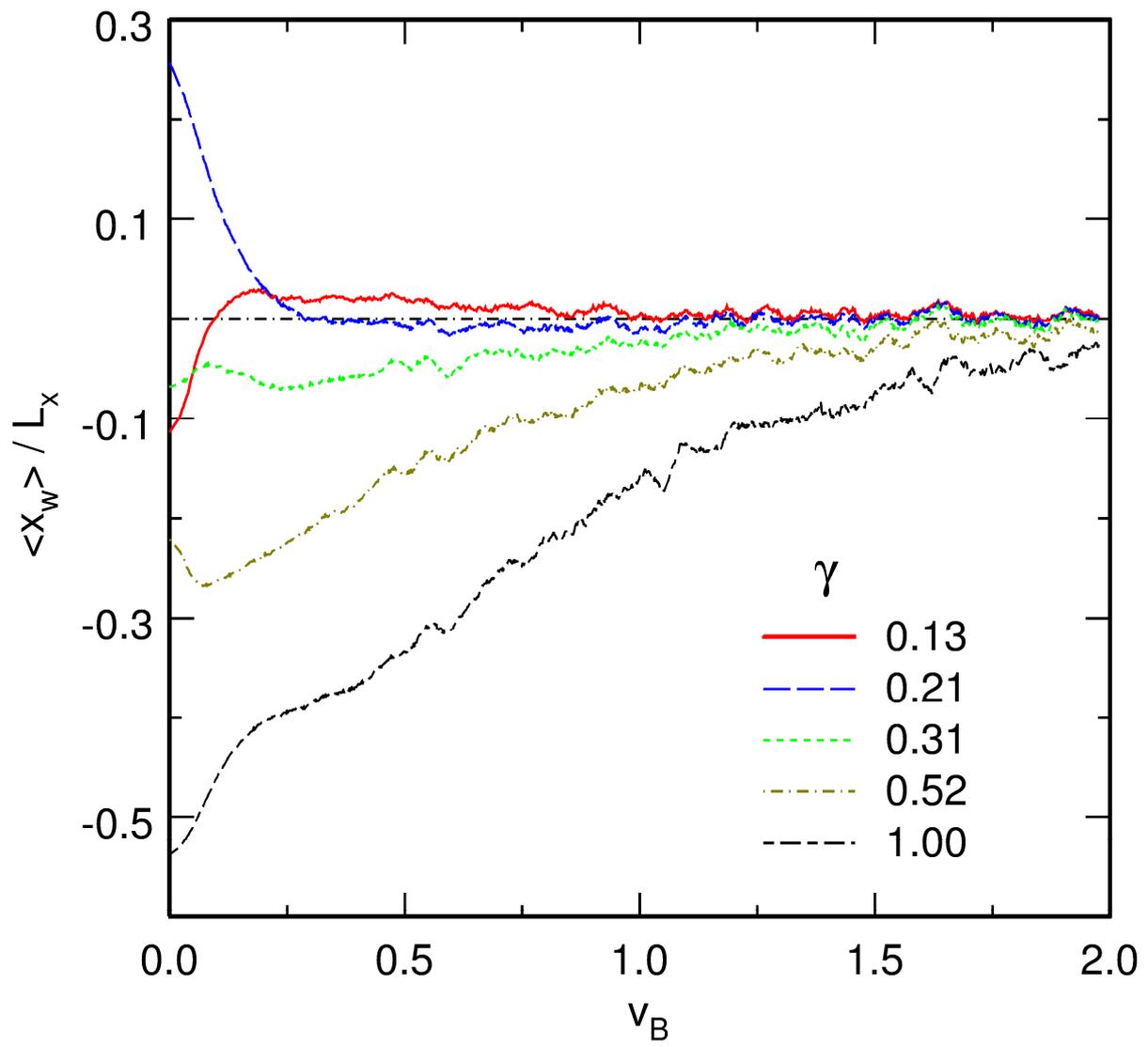

**Figure 7**

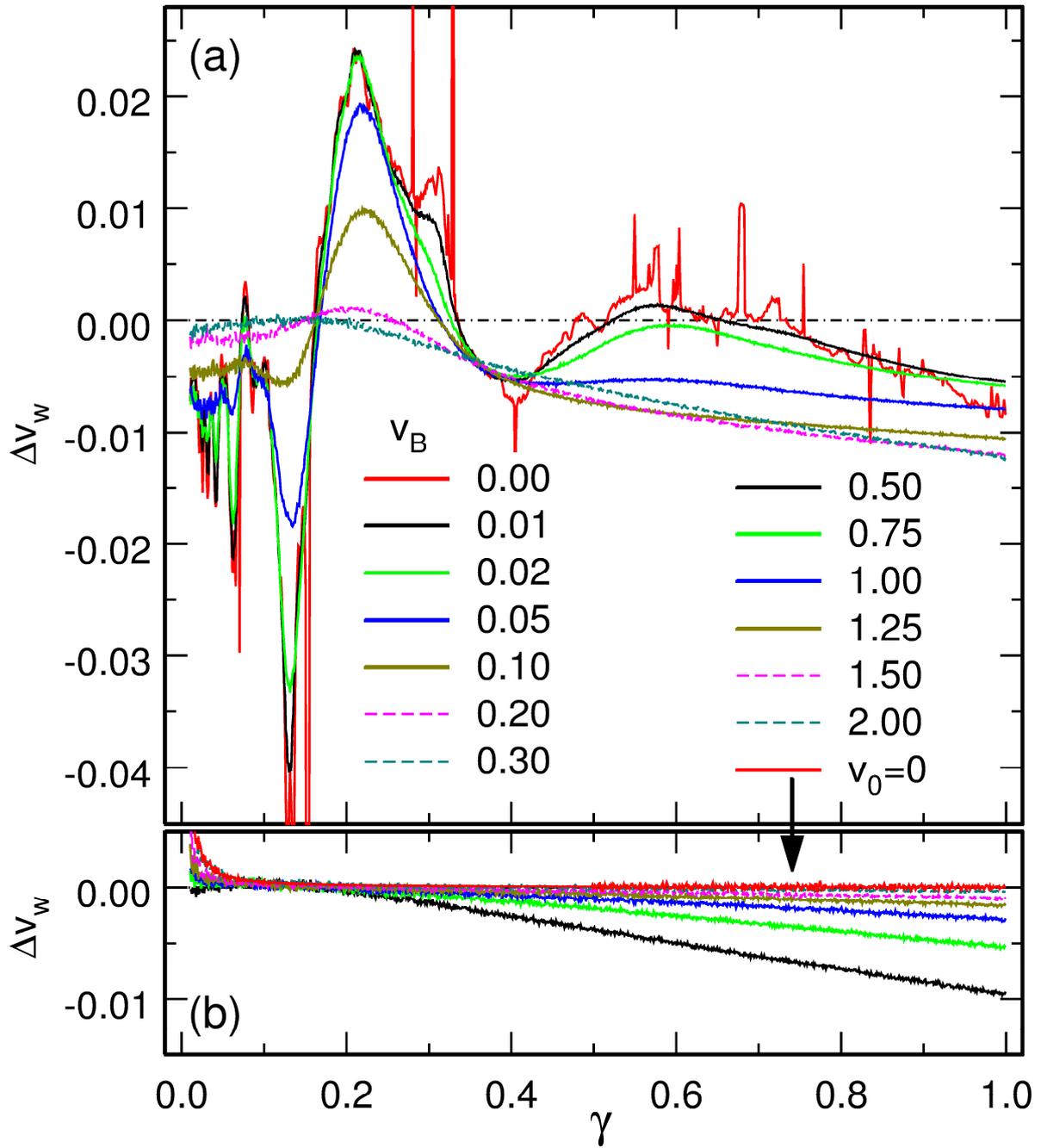

**Figure 8**

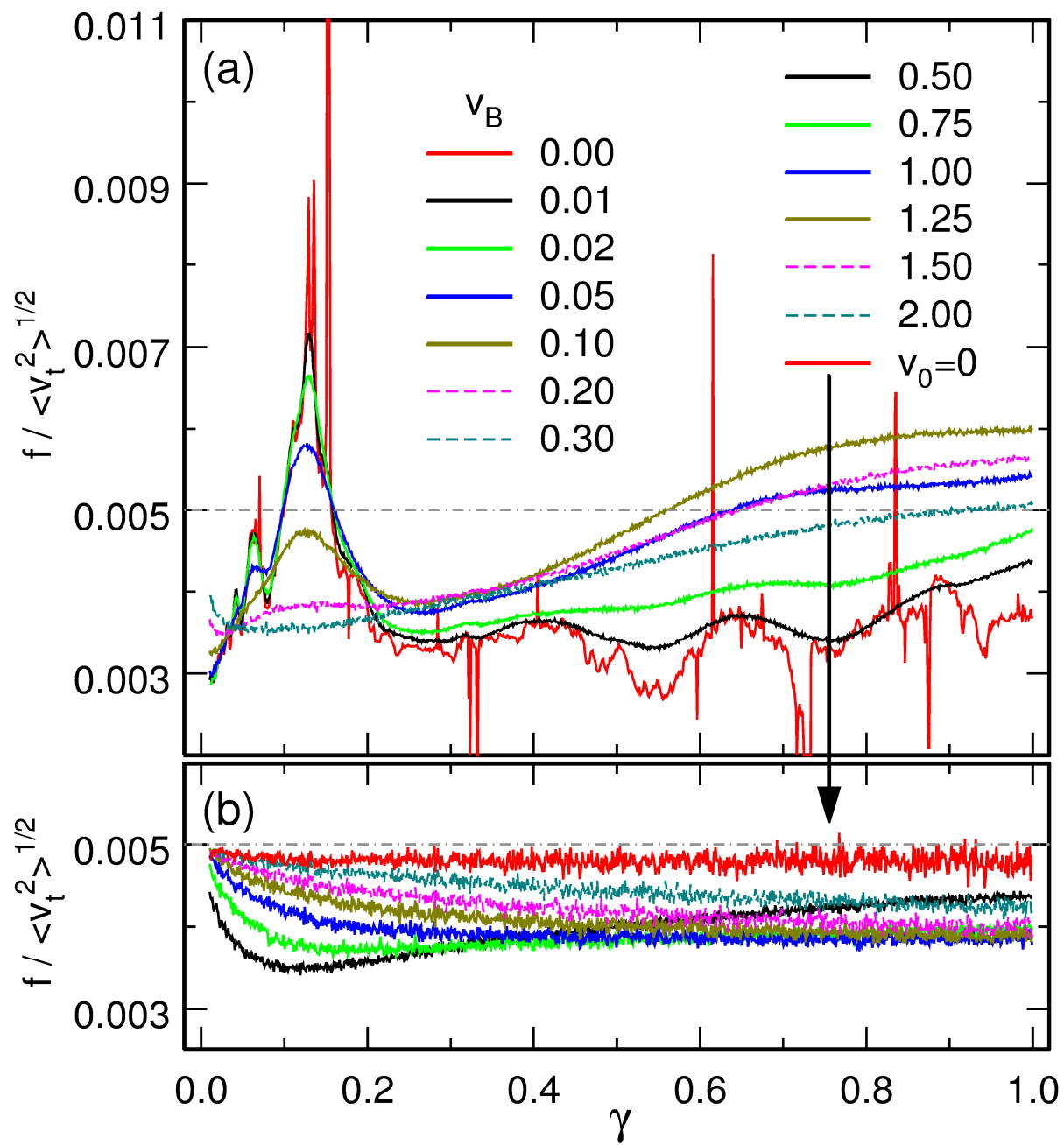

**Figure 9**

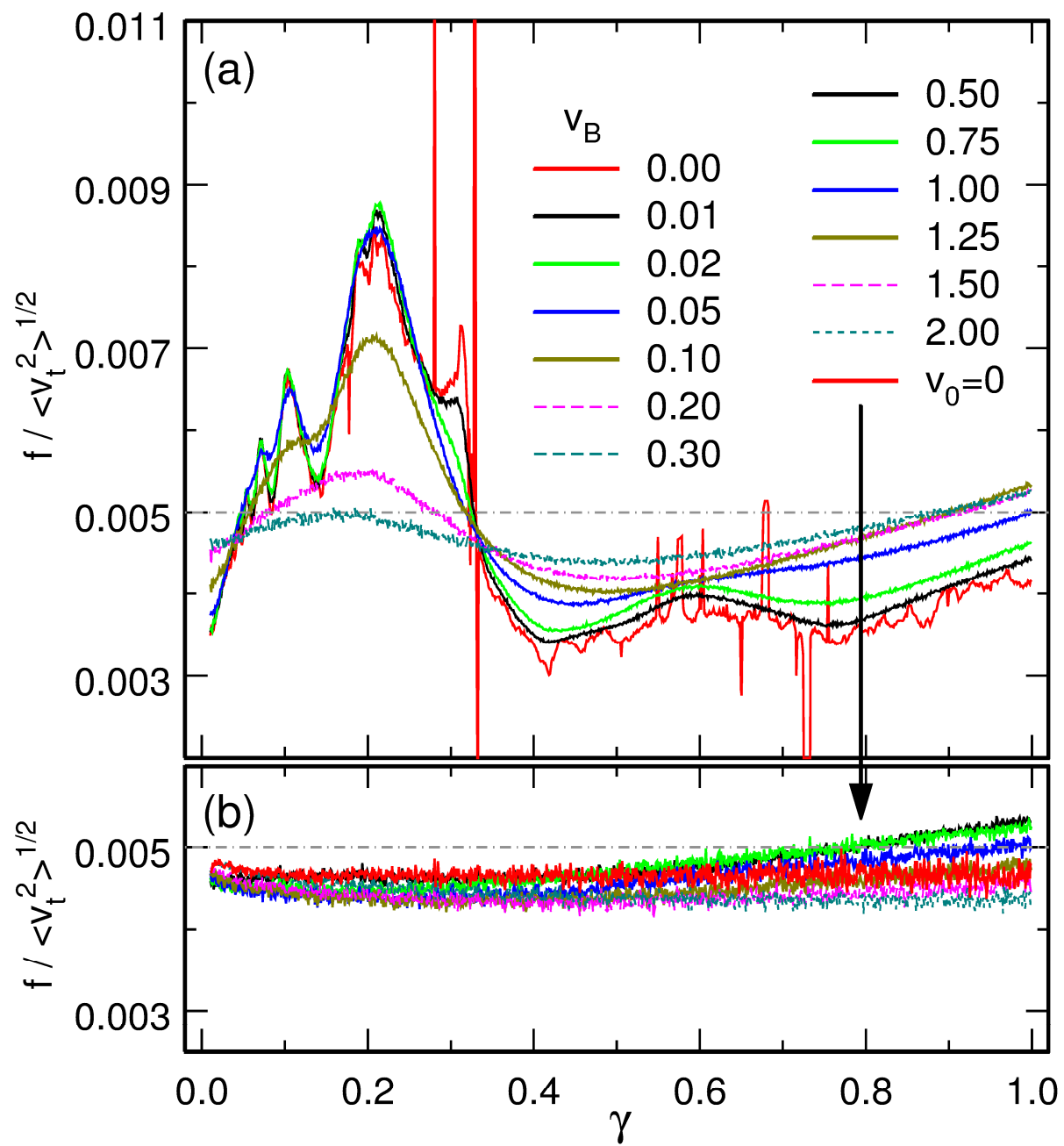

**Figure 10**

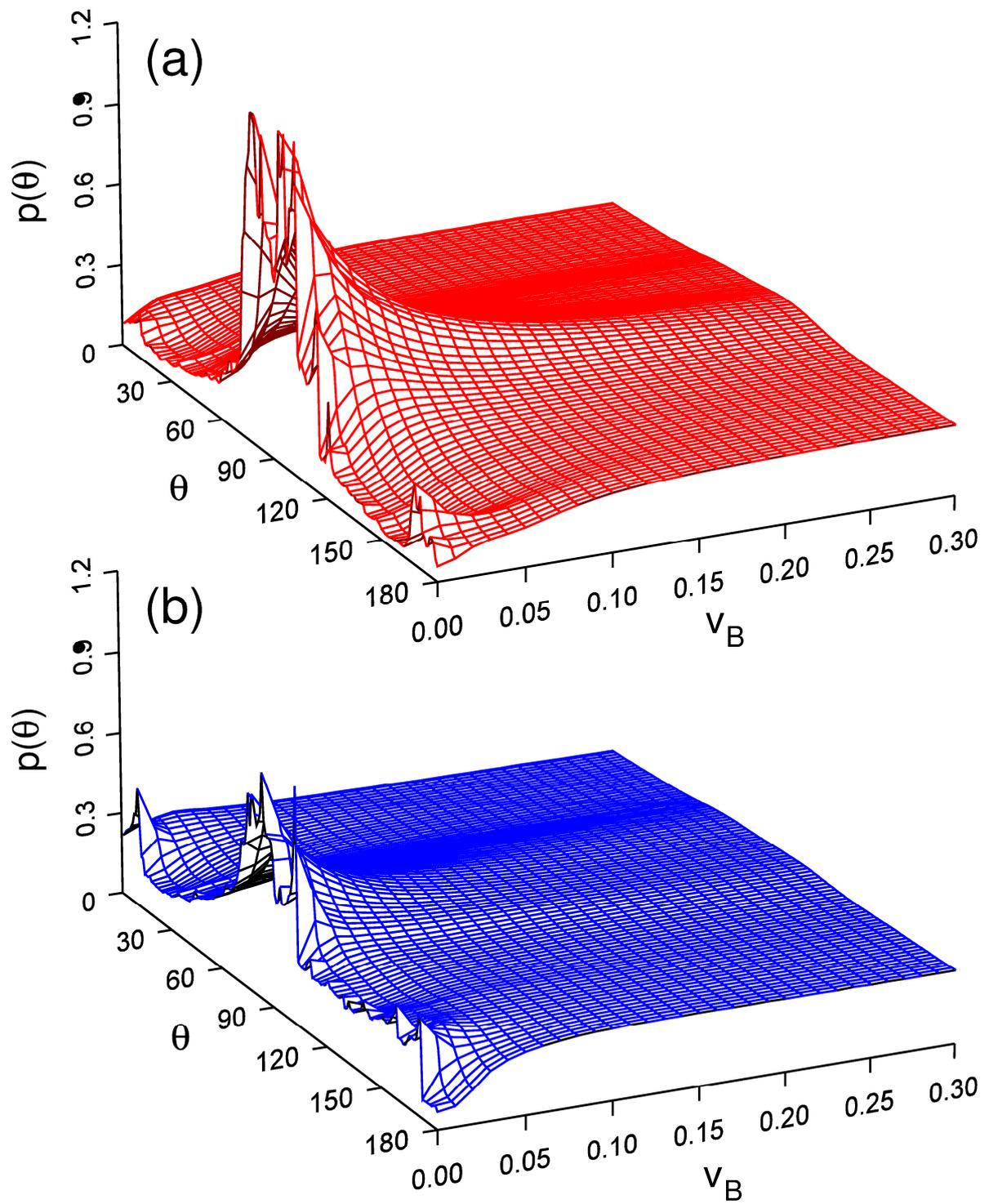

**Figure 11**

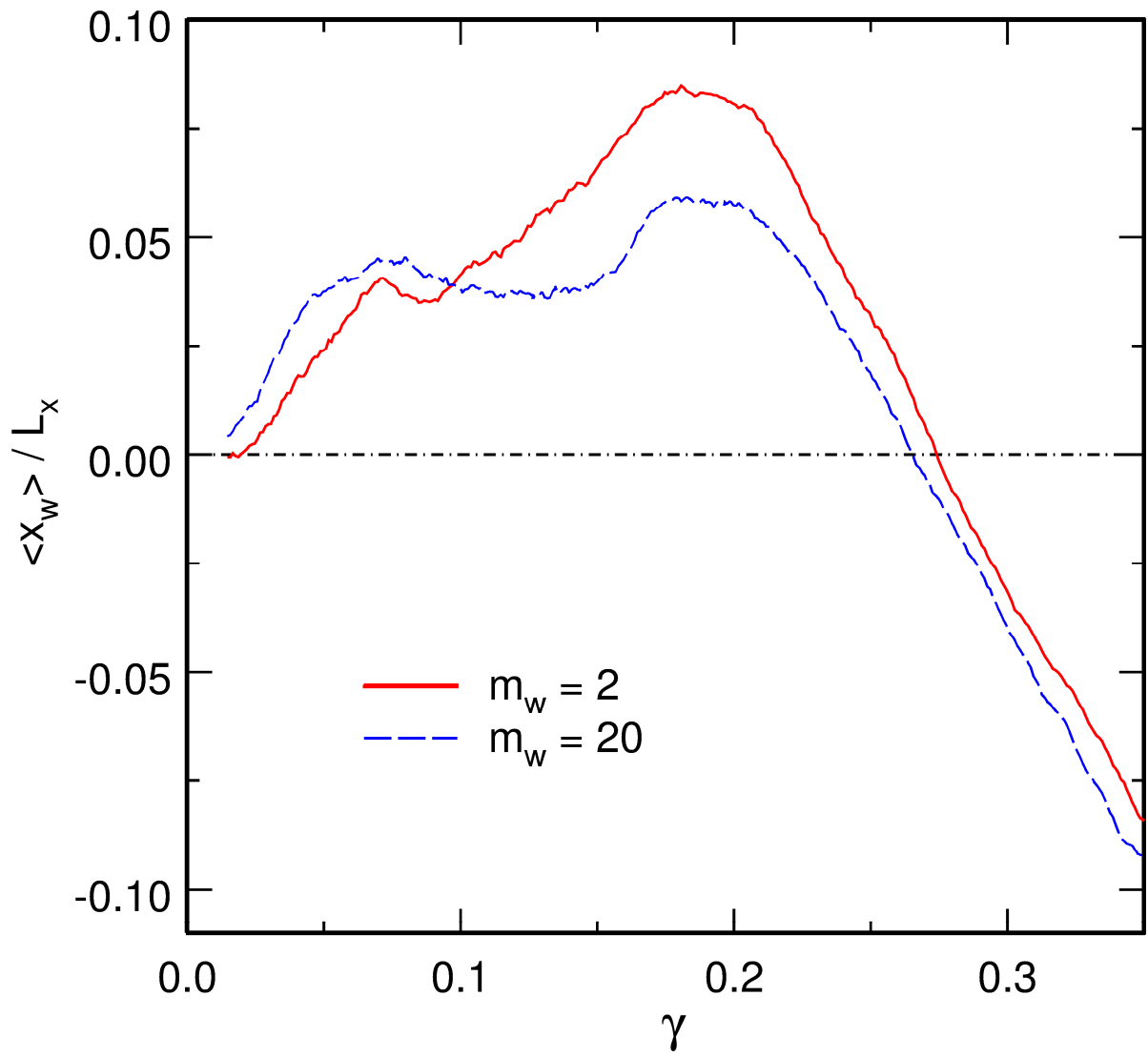

**Figure 12**

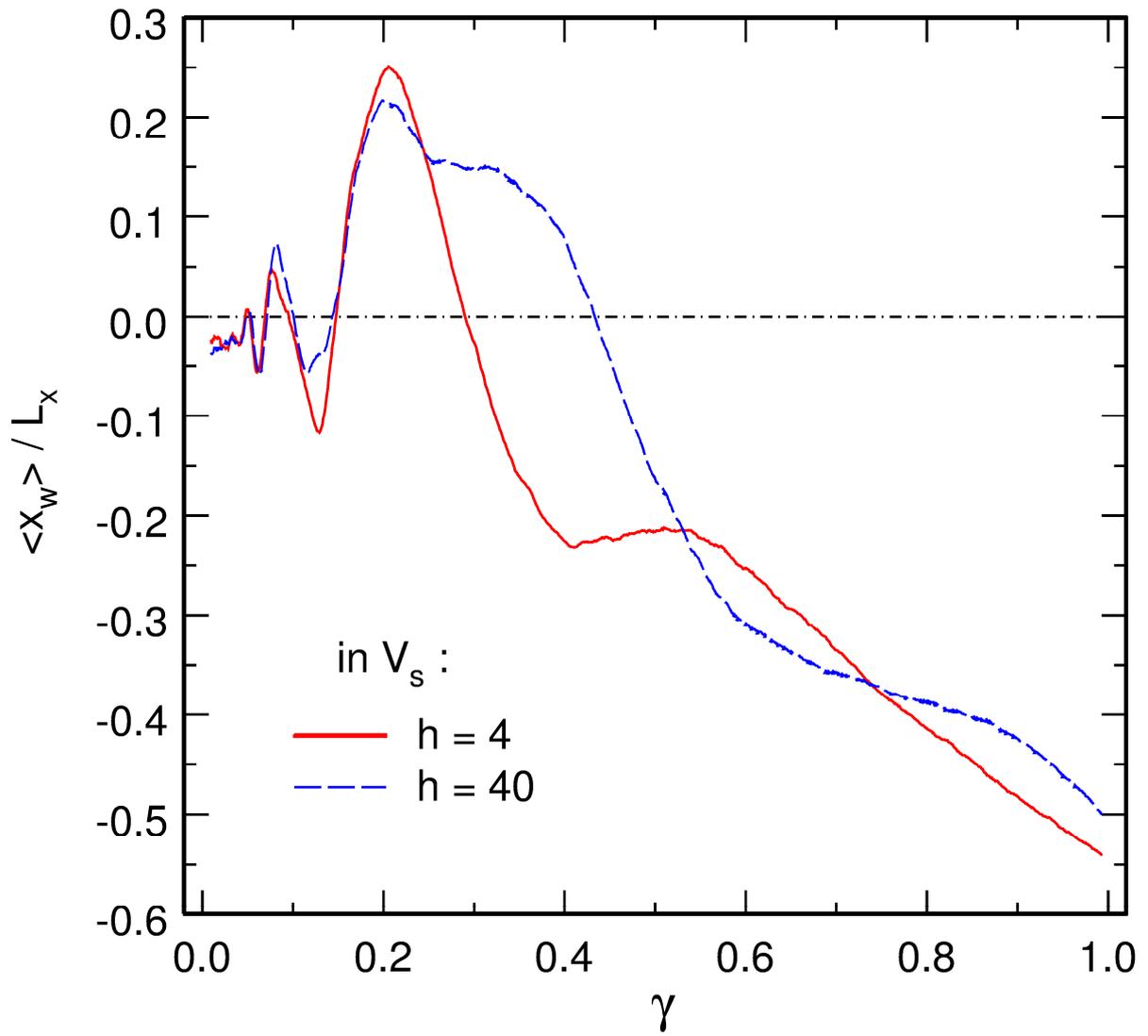

**Figure 13**

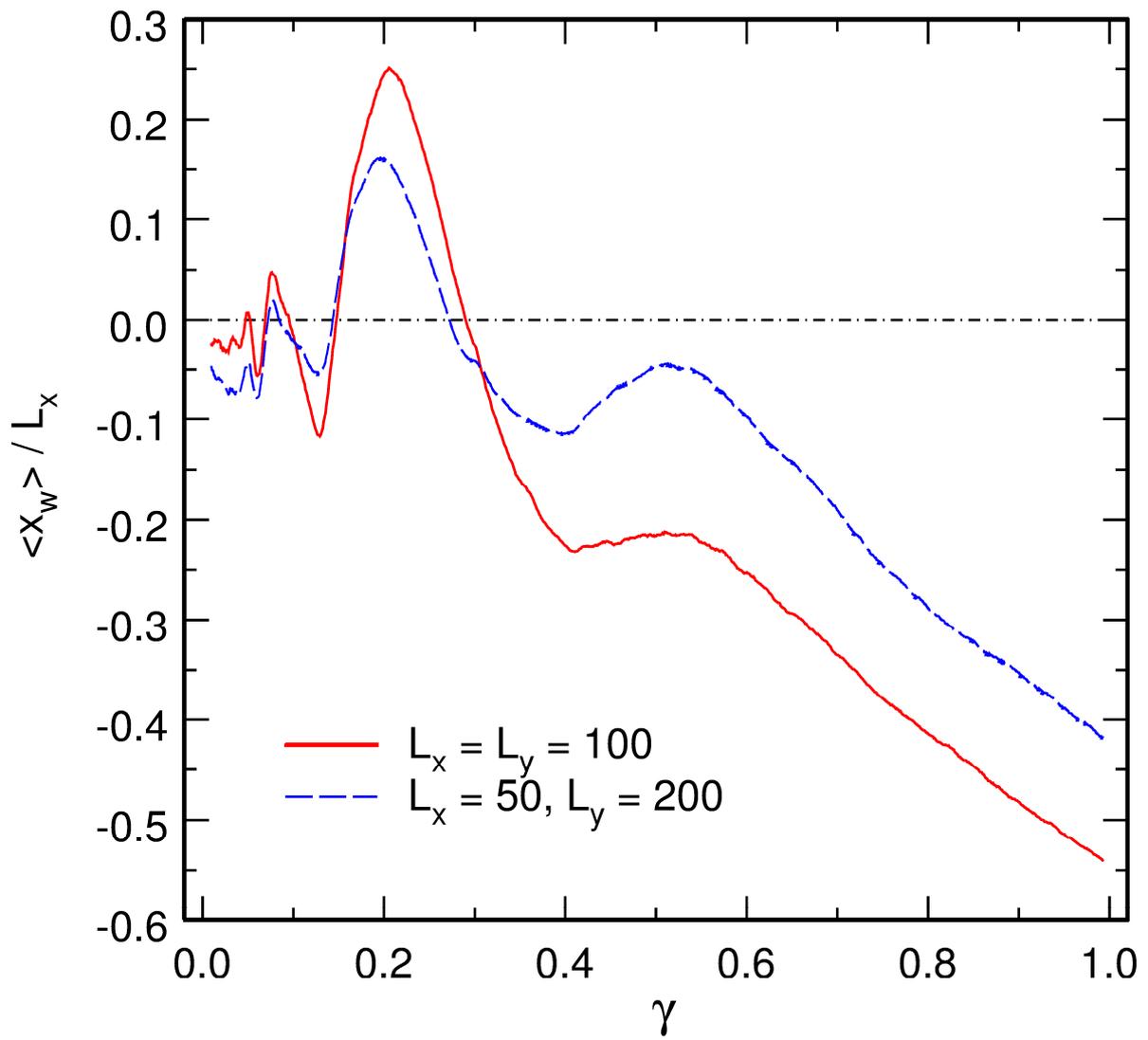

**Figure 14**

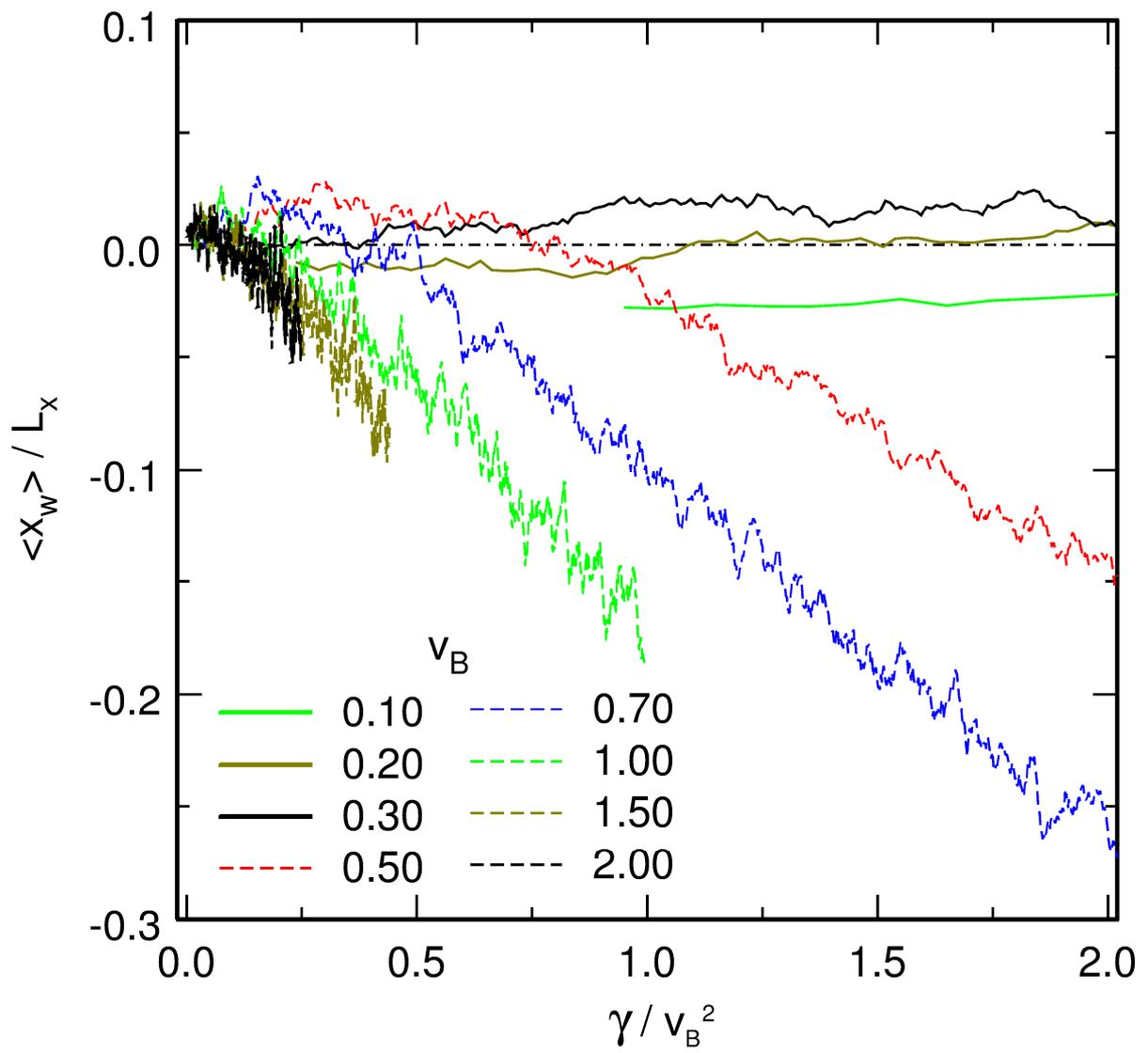